\documentclass[aps,prl,amsmath,reprint,superscriptaddress]{revtex4-1}
\usepackage{bm}
\usepackage{graphicx}
\usepackage{float}
\usepackage{color,soul}
\usepackage[dvipsnames]{xcolor}
\usepackage[colorlinks=true, linkcolor=blue, citecolor=blue, urlcolor=blue,breaklinks=true]{hyperref}
\newcommand{\ket}[1]{|#1\rangle}             % Ket Dirac's notation %
\newcommand{\bra}[1]{\langle#1|}             % Bra Dirac's notation  %
\begin{document}
	
	\title{Quantum steering with vector vortex photon states\\ with the detection loophole closed}
	
	\author{Sergei Slussarenko}
	\email{s.slussarenko@griffith.edu.au}
	\author{Dominick J. Joch}
	\affiliation{Centre for Quantum Dynamics and Centre for Quantum Computation and Communication Technology, Griffith University, Brisbane, Queensland 4111, Australia}
	\affiliation{These authors contributed equally}
	\author{Nora Tischler}
	\affiliation{Centre for Quantum Dynamics and Centre for Quantum Computation and Communication Technology, Griffith University, Brisbane, Queensland 4111, Australia}
	\affiliation{ Dahlem Center for Complex Quantum Systems, Freie Universit\"{a}t Berlin, 14195 Berlin, Germany }
	\author{Farzad Ghafari}
	\affiliation{Centre for Quantum Dynamics and Centre for Quantum Computation and Communication Technology, Griffith University, Brisbane, Queensland 4111, Australia}
	\author{Lynden K. Shalm}
	\author{Varun B. Verma}
	\author{Sae Woo Nam}
	\affiliation{National Institute of Standards and Technology, 325 Broadway, Boulder, Colorado 80305, USA.}
	\author{Geoff J. Pryde}
	\email{g.pryde@griffith.edu.au}
	\affiliation{Centre for Quantum Dynamics and Centre for Quantum Computation and Communication Technology, Griffith University, Brisbane, Queensland 4111, Australia}

	\begin{abstract}
		Violating a nonlocality inequality enables the most powerful remote quantum information tasks and fundamental tests of quantum physics. Loophole-free photonic verification of nonlocality has been achieved with polarization-entangled photon pairs, but not with  states entangled in other degrees of freedom. Here we demonstrate completion of the quantum steering nonlocality task, with the detection loophole closed, when entanglement is distributed by transmitting a photon in an optical vector vortex state, formed by optical orbital angular momentum (OAM) and polarization. As well as opening up a high-efficiency encoding beyond polarization, the critically-important demonstration of vector vortex steering opens the door to new free-space and satellite-based secure quantum communication devices and device-independent protocols. 
		
	\end{abstract}
	\maketitle
	\section{Introduction}
	Optical quantum correlations are essential for quantum information science, with applications ranging from computation to metrology and communication~\cite{rev_flamini18, slussarenko17n, rev_slussarenko19}. Quantum nonlocality, tested via e.g.\  violation of a Bell inequality~\cite{hensen15,giustina15,shalm15,bbt18,rauch18} or a quantum steering inequality~\cite{wittmann12}, enables device-independent-class quantum communication protocols~\cite{acin06,branciard12} and certified randomness generation~\cite{law14,bierhorst18}. Such a rigorous verification requires the closure of various loopholes~\cite{larsson14} that appear when there are assumptions that may not hold and that allow classical physics or an eavesdropper to simulate nonlocality when there is none. Experimentally, the most challenging loophole to close is the detection loophole, associated with the fair sampling assumption~\cite{bennet12}. If photon transmission and detection efficiencies are not high enough, then the statistics of correlations may be spoofed by hiding cheating strategies as apparent loss. 
	
	To date, loophole-free observation of quantum nonlocality with photons has been limited to polarization qubits~\cite{giustina15,shalm15,wittmann12}. In particular, the detection loophole has not been closed in experiments encoded in other photon degrees of freedom, which include transverse spatial~\cite{rev_erhard18} (including orbital angular momentum, OAM), temporal~\cite{ansari18} and frequency~\cite{cai17} modes. For some photonic quantum information tasks, polarization encoding is known to be non-optimal~\cite{korzh15,ecker19,zhu21}.  Using a different degree of freedom can provide advantages particular to the specific physical encoding---such as improved noise tolerance. One can either encode two-dimensional qubits or, if advantageous, encode higher-dimensional qudits to improve the information carrying capacity of photons. 
	
	 Here we show a detection-loophole-free demonstration of a photon-encoded non-locality test that involves a degree of freedom other than polarization. In this proof-of-principle experiment, we have chosen a two-dimensional vector vortex encoding, which has the advantage that it  avoids the need for  rotational  alignment between the reference frames of the two parties that share the entangled state~\cite{aolita07,dambrosio12natcomm,vallone14} via the transmission of the photon so encoded.  Rotation invariance is beneficial for applications where quantum information needs to be transferred via a free-space channel to a portable receiver, e.g.\ near- or mid-distance communications to moving aerial receivers, or satellite communication. Alternatively, vector vortex modes could assist in quantum channel characterization for error-tolerant applications~\cite{ndagano17} or allow for encoding high-dimensional qudits, but this would then require giving up the feature of rotation-invariance.   
	
	 As the nonlocality test we perform  quantum (or Einstein-Podolsky-Rosen)  steering~\cite{wiseman07,rev_cavalcanti16,rev_uola20}  of qubits, where one of the photon qubits is encoded in vector vortex modes formed by a combination of OAM and polarization states of light.   Quantum steering is a nonlocality test in which one party trusts quantum mechanics to describe its measurement apparatus, but no assumptions are made on the other party and the source of nonlocal states.  
	
	Closing the detection loophole in a nonlocality test requires an entangled state that is both highly correlated and is distributed and detected with high efficiency (low loss). The use of steering protocols can reduce the stringent requirements for the efficiency in transmitting the state to the untrusted party, compared to a Bell test, making these tests more suitable for environmental conditions with higher amounts of noise and loss. 
	However, the majority of current methods to generate, measure or transfer information in degrees of freedom other than polarization are limited in efficiency or quality and cannot provide performance sufficient for a loophole-free nonlocality test.  Few examples include spatial light modulators for spatial mode control, or integrated and free-space electrooptic elements for time and frequency control that provide flexible manipulation of the relevant degree of freedom but introduce a significant amount of loss. In pursuing the use of different degrees of freedom in entanglement verification, recent works address additional loopholes relevant to their specific type of encoding~\cite{vedovato18}, or aim at substantially improving the state quality and measurement efficiency of high-dimensional states~\cite{valencia20}. Nevertheless,  despite significant progress in complex experimental demonstrations involving different physical encodings of quantum information~\cite{malik16,zhang17n,kues17,erhard18,wang18,reimer19}, their application to highly-demanding protocols such as device-independent communication remains a challenge. 
	
	The key elements  in our solution to overcome this roadblock  are: that a polarization-encoded photon, entangled with another such photon, can be efficiently converted so that its information is stored in a rotationally-invariant vector vortex state; that this information can be then collected by a (potentially rotated) observer; and that the photon can be subsequently measured using efficient projective measurement based on a mode conversion element, polarization optics and single-photon detection (Fig.~{\bf 1a}). It is important that all of this must be achieved while keeping the quality of the quantum correlations very high. By using state-of-the-art generation, mode conversion and detection of photons we simultaneously achieved the goals of high entangled-state quality and high quantum-channel transmission and thus demonstrated detection-loophole-free nonlocal correlations, with and without rotated observers.
	\begin{figure*}
		\includegraphics{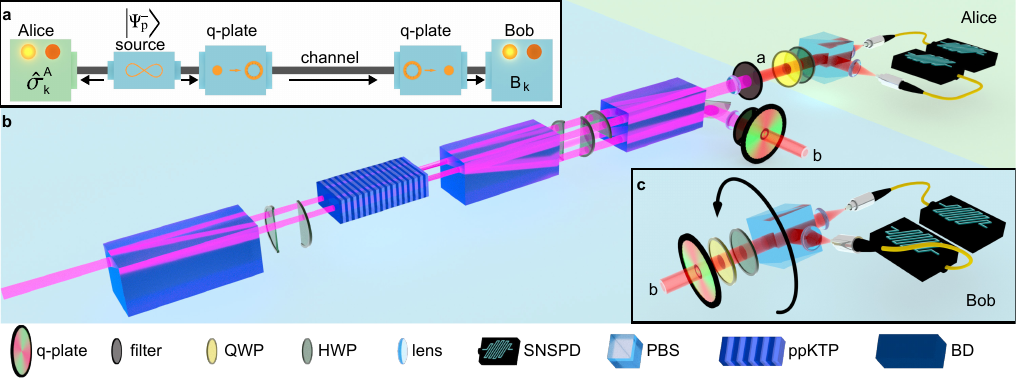}
		\caption{ \label{fig:apparatus} {\bf Experimental concept and setup.} {\bf a} Conceptual diagram of vector vortex steering. {\bf b} Experimental setup. A photon-pair source~\cite{shalm15,tischler18p} produces a polarization-entangled singlet state. The polarization qubit of one photon is converted to a vector vortex state qubit through a q-plate and is sent through a free-space channel (4F imaging system not shown) to Bob. Alice's measurement apparatus performs a two-outcome projective measurement on a polarization qubit. {\bf c} Bob's measurement apparatus has an additional q-plate that converts a vector vortex state qubit into a polarization qubit, and is mounted in a cage system in order to allow for a rotation around the beam propagation axis, as described in the main text. Blue and green backgrounds represent untrusted and trusted devices, respectively.}
	\end{figure*}
	\section{Results}
	\subsection{Protocol}
	We first summarize the key elements of our test, namely the encoding method for the transmitted qubit, the quantum steering test and the technology that enabled it. Our logical vector vortex qubit is encoded with the zero  total angular momentum basis states
	\begin{equation}\label{states}
		\ket{0}=\ket{L,l=-1}; \ket{1}=\ket{R,l=1}.
	\end{equation}
	Here $\ket{L}$ and $\ket{R}$ denote left- and right-circular polarization states, and $\ket{l=\pm1}$ denote states with $\pm\hbar$ of OAM per photon. 
	
	Quantum steering is a type of nonlocal correlation that is strictly weaker than Bell nonlocality, but stronger than entanglement~\cite{wiseman07}. Unlike Bell nonlocality, steering is an asymmetric protocol~\cite{tischler18p}. One of the parties, Alice, is trusted: her measurement process is described by quantum mechanics and her measurement outcomes are accurate and are honestly communicated. This assumption allows performing correlation tests that are more resilient to noise~\cite{saunders10} and loss~\cite{bennet12} in the untrusted channel. Steering is obviously interesting for protocols where one party is in a secure location (e.g.\ at home base) while another (e.g.\ in the field) may be more susceptible to the  action of an adversary. (However, by replacing classical with quantum instructions, the need for trust in one party can be significantly reduced~\cite{kocsis15}.)  These include one-side device-independent quantum key distribution~\cite{branciard12} and randomness certification~\cite{passaro15} protocols, and any other application of quantum nonlocality to e.g. device independence where one of the parties may be considered trusted because of their secure location~\cite{rev_cavalcanti16,rev_uola20}.  Besides polarization qubits, quantum steering has been demonstrated with high-dimensional~\cite{zeng18,dabrowski18, designolle20} and single-photon path-entangled~\cite{guerreiro16} quantum states, although a loophole-free demonstration of high-dimensional steering remains a challenge.
	
	In our test, we used the loss-tolerant steering inequality introduced in Ref.~[\onlinecite{bennet12}]. In this protocol, upon receiving one of the particles, Alice chooses a particular measurement setting $k$ from a predetermined set of $n$ measurements to perform on the particle and announces the same measurement setting to Bob. She then performs her measurement $\hat{\sigma}^A_k$ and compares the outcome with the result $B_k$, which she obtained from Bob. The process is repeated to accumulate statistics, and quantum steering is verified if the observed correlations 
	\begin{equation}\label{steering}
		S_n= \frac{1}{n} \sum_{k=1}^{n}\langle\hat{\sigma}^A_k{B_k}\rangle
	\end{equation}
	are higher than a corresponding bound $C_n$.  In order to close the detection loophole, the untrusted party, Bob, is required to announce his measurement outcome a specific minimum fraction of times~\cite{bennet12}, setting a stringent requirement on the transmission  $\xi$ of the channel between the source and Bob. The bound is thus also dependent on $\xi$, $C_n=C_n(\xi)$.  In contrast, because Alice is trusted, no transmission requirement is set on the transmission between Alice and the source. A randomized measurement choice and particular requirements on the locations of Alice and Bob and the timing of their communications allow closure of the locality and the freedom of choice loopholes~\cite{wittmann12}. Closing these loopholes was not a goal of this study because doing so is independent of the state encoding. 
	\subsection{Experiment}
	Our experimental apparatus consisted of a high-heralding-efficiency entangled photon pair source, mode conversion (polarization to vector vortex state) and reverse conversion optics, polarization measurement optics and high-efficiency single photon detection (Fig.~{\bf 1b}). We used a group-velocity-matched~\cite{weston16} spontaneous parametric downconversion (SPDC) process to generate telecom-wavelength polarization-entangled photon pairs. A nonlinear periodically-poled potassium titanyl phosphate (ppKTP) crystal was embedded into a beam displacer (BD) interferometric setup~\cite{evans10,shalm15,bierhorst18,tischler18p}, as shown in Fig.~{\bf 1b} to produce a state having high fidelity with the polarization-encoded singlet Bell state 
	
	\begin{equation}\label{singlet}
		\ket{\Psi^-_{\rm p}}=\frac{1}{\sqrt2}(\ket{H}_{\rm a}\ket{V}_{\rm b}-\ket{V}_{\rm a}\ket{H}_{\rm b}),
	\end{equation}
	with $\ket{H}$ and $\ket{V}$ denoting horizontal and vertical modes, respectively. Indices ``a'' and ``b'' label the photons that were then sent to Alice and Bob, respectively. Combined with the use of high-efficiency superconducting nanowire single-photon detectors~\cite{marsili13} (SNSPDs) this source architecture has shown the capability to achieve very high heralding (Klyshko~\cite{klyshko80}) efficiencies~\cite{shalm15,bierhorst18}. At the same time, BD-based sources have shown the ability to generate high-quality entanglement in the polarization degree of freedom, with experimentally obtained states $\rho_{\rm exp}$ with fidelities up to $\mathcal{F}_{\rm p}={\bra{\Psi^-_{\rm p}}}\rho_{\rm exp}\ket{\Psi^-_{\rm p}}\geq 0.997$ with the two-photon singlet state $\ket{\Psi^-_{\rm p}}$~\cite{shalm15,bierhorst18,tischler18p}.
	
	The key challenge in using spatial modes for encoding the entangled qubits it is to maintain high transmission efficiency over a complete optical path while achieving high mode conversion fidelity. This requirement currently rules out most common methods such as the ones based on spatial light modulators, digital mircomirror devices, or metasurfaces due to their limited diffraction efficiency. In our experiment  we used q-plates~\cite{marrucci06,slussarenko11,rev_marrucci11,rev_rubano19} to transform polarization qubits into vector vortex mode qubits. Q-plates (and geometric-phase-based mode transformation elements in general~\cite{bhandari97}) induce a polarization transformation that varies with the coordinate in the plane perpendicular to the beam propagation. This reshapes the wavefront with a non-uniform geometric phase. A q-plate acts on an input beam by reversing its circular polarization component and adding an OAM value of $\pm2q\hbar$ (``$+$'' for input $\ket{L}$ and ``$-$'' for input $\ket{R}$ polarization), with $q$ being the topological charge of the q-plate pattern. A q-plate with $q=0.5$ directly converts~\cite{cardano12} a polarization qubit into a rotation-invariant vector vortex mode qubit~\cite{dambrosio12natcomm}.  We use devices based on patterned soft matter~\cite{marrucci06a,tabiryan09}, that to date are the only ones that have shown the capability to perform high-quality transformation with virtually no loss, although a number of other q-plate platforms also exist~\cite{bomzon01a,karimi14,rev_piccardo20}.
	
	Our experimental setup is shown in Fig.~{\bf 1b}-{\bf c}. Photon pairs in the $\ket{\Psi^-_{\rm p}}$ state were generated by the collinear telecom SPDC source. The downconverter was pumped by a $1$ ps-pulse-length mode-locked Ti:Sapphire laser. We used an antireflection-coated silicon filter to separate the pump from the downconverted photons. The two-photon coincidence rate was $\sim 20,000$ s$^{-1}$. The photons were coupled into optical fibers (not shown in Fig.~{\bf 1}, for simplicity) before emerging into free space for transmission and measurement. Photon $a$ was sent to the trusted party, Alice, where she performed a polarization projection measurement with a quarter-wave plate (QWP), half-wave plate (HWP), and polarizing beam splitter (PBS). No degree of freedom conversion was performed on this qubit because no minimum requirement is set on the transmission between source and trusted party, making the use of any degree of freedom trivial on this channel. Photon $b$ was sent through a q-plate where its qubit was converted from polarization to vector vortex mode degree of freedom, providing the overall two-photon shared state~\cite{cozzolino19}
	\begin{equation}
		\ket{\Psi^-}=\frac{1}{\sqrt2}(\ket{H}_{\rm a}\ket{1}_{\rm b}-\ket{V}_{\rm a}\ket{0}_{\rm b}).
	\end{equation} 
	The photon was then sent through a free-space channel, which consisted of a 4F imaging system to the analyzer device of the untrusted party, Bob. Bob's measurement device consisted of another $q=0.5$ q-plate,  followed by a QWP, HWP and PBS. This allowed analyzing vector vortex modes by converting them via polarization to path, where the photon could finally be fiber-coupled and detected.  All of Bob's optical components (Fig.~{\bf 1c}), except the detectors themselves, were mounted in a cage system that could rotate around the propagation axis of the photons.  This effectively amounted to a rotation of the state from Bob's point of view.  We refer to the rotation setting of Bob's apparatus as its \textit{orientation}. After the PBS, the photons were collected into single-mode telecom fibers and sent to SNSPDs for readout. All optical components, including the q-plates, were anti-reflection coated for $1550~{\rm nm}$.  The use of a 4F imaging system was required because q-plates, and similar devices that perform a phase-only transformation, do not generate a propagation-invariant optical mode, such as a Laguerre-Gauss mode with single radial and azimuthal numbers. This would result in a non-Gaussian mode shape after the reverse transformation at Bob's stage, and thus reduced fiber coupling efficiency. A 4F setup can compensate for this effect on laboratory-scale distances. 
	\begin{figure}
		\includegraphics{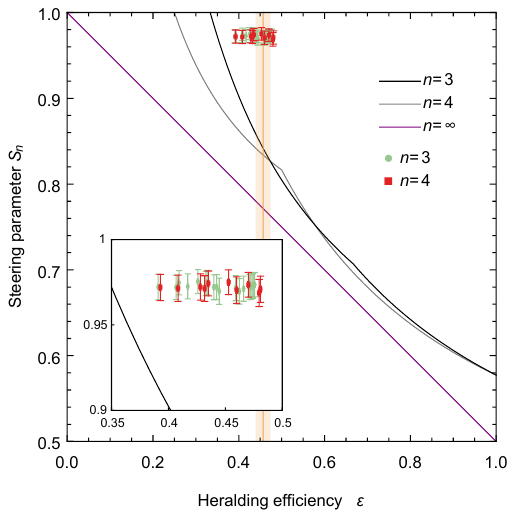}
		\caption{ \label{fig2} {\bf Experimental vector vortex quantum steering.} Lines represent steering bounds for different numbers of measurement settings. Green circles and red squares correspond to the steering parameters observed at different orientation angles of Bob's receiver, measured with $n=3$ and $n=4$ measurement settings, respectively. Error bars and shaded areas correspond to a measurement uncertainty equal to two standard deviations total. The systematic uncertainty in steering parameter values was evaluated according to Ref.~[\onlinecite{bennet12}]. The orange vertical line denoting the average heralding efficiency observed with polarization-only steering is shown for comparison.}
	\end{figure}
	\begin{figure}
		\includegraphics{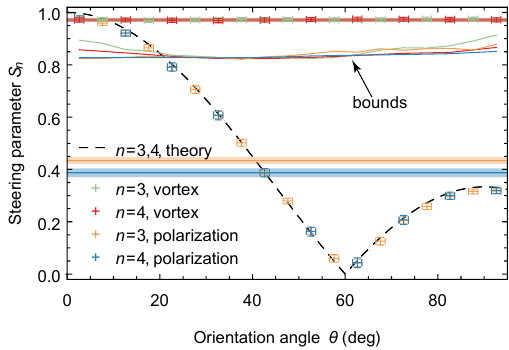}
		\caption{ \label{fig3} {\bf Quantum steering as a function of the orientation angle of Bob's measurement apparatus.}  Dots are the experimentally measured data, and the lines with corresponding colors without shaded areas represent steering bounds calculated for each point, taking into account the number of measurement settings and measured heralding efficiency. Lines with shaded areas correspond to the dynamic steering scenario when $\theta$ was adjusted during the measurement, see main text for details. The dashed black line is the theoretical prediction for steering with polarization qubits. Error bars and shaded regions correspond to a measurement uncertainty equal to two standard deviations total.}
	\end{figure}
	
	With our setup, we could change between the hybrid $\ket{\Psi^-}$ and polarization-only $\ket{\Psi^-_{\rm p}}$ encodings by removing or adding back the q-plates in Bob's arm. First, we characterized the general quality of the polarization-entangled states and the channel transmission. We analyzed the distributed states using quantum state tomography~\cite{james01}. The maximum fidelity of the experimentally generated state with the ideal state $\ket{\Psi^-_{\rm p}}$ was determined to be $\mathcal{F}_{\rm p}=0.982\pm0.001$. As the orientation of Bob’s apparatus was changed, the reconstructed state deviated from $\ket{\Psi^-_{\rm p}}$, and so the fidelity fell from the maximum value to $\mathcal{F}_{\rm p}=0.001\pm0.004$ over rotation angles of $\theta=0^{\circ}$ to $\theta=90^{\circ}$, as expected. The purity of the state hardly varied, and we determined an average value of $P_{\rm p}=0.970\pm0.002$. Over the same orientation range of Bob’s apparatus, we observed Bob's heralding efficiencies ranging from $\varepsilon_{\rm p}^{\rm min}=0.421\pm0.001$ to $\varepsilon_{\rm p}^{\rm max}=0.479\pm0.001$ (average $\bar\varepsilon_{\rm p}=0.46\pm0.02$) due to the small hardware imperfections discussed below. We repeated these measurements for the entangled state $\ket{\Psi^-}$ with vortex vector encoding, and determined that the entangled state fidelity hardly varied with the orientation of Bob’s apparatus, with an average value of $\mathcal{F}=0.977\pm0.003$. Similarly, the purity stayed nearly constant, with an average value $P=0.975\pm 0.002$. With the same rotation as above, Bob’s heralding efficiencies ranged from $\varepsilon^{\rm min}=0.391\pm 0.001$ to $\varepsilon^{\rm max}=0.481\pm0.001$ (average $\bar\varepsilon=0.45\pm0.03$). These results indicated that quantum steering with vector vortex qubits should be possible with the detection loophole closed, even when as few as $n=3$ measurement choices~\cite{bennet12} are used in the protocol.
	
	Next, we attempted quantum steering with the vector vortex encoded state. Our results with $n=3$ and $n=4$ measurement settings are shown in Fig.~{\bf 2}, where each of the data points represents a different angular orientation of Bob’s receiver. The strength of the correlations and heralding efficiency were sufficient for the steering parameter to remain above the steering bound for the entire range of receiver orientation angles. The angular distribution is presented in Fig.~{\bf 3}, where it can be seen that, by comparison, the polarization-encoded state has a steering parameter that varied strongly and falls beneath the band for a large range of receiver orientations. 
	
	In addition, we performed a set of quantum steering measurements in which we simulated a dynamically rotating receiver. For each measurement setting, Bob's receiver apparatus was oriented at a different angle, sampling uniformly in the range from $\theta=0^{\circ}$ to $\theta=90^{\circ}$, resulting in an averaging effect. The results are also shown in Fig.~{\bf 3}, where it can be seen that the vector vortex state allows steering to be strongly demonstrated in this noisy scenario, whereas the polarization-encoded state does not.
	
	\section{Discussion}
	In summary, we have shown that it is possible to use a hybrid polarization-OAM degree of freedom to distribute nonlocal correlations with the detection loophole closed. Specifically, the conversion, transmission, and reverse conversion maintained a high state quality and efficiency, enabling a postselection-free violation of a quantum steering inequality in the presence of loss, with $n=3$ measurement settings. Among the observed experimental imperfections, we attribute the non-unit purity of the distributed state to the low contrast of Bob's PBS, which, being the central part of the cage system, could not be fine-adjusted by tilt or rotation. We attribute the loss in efficiency to the lack of rigidity of Bob's rotating receiver, which was deforming under its own weight as the orientation was varied. We verified this assumption using a separate  transmission setup with a stationary version of Bob's receiver apparatus. We observed less than $5\%$ increase in loss (including an increase in fiber coupling loss), when vector vortex conversion, 4F imaging, and reverse conversion optical elements were added, compared to a propagation and fiber coupling of a Gaussian beam. The observed negligible loss due to conversion to and from the vector vortex mode suggests that the heralding efficiency required for Bell tests or one-side device-independent quantum key distribution~\cite{branciard12} can be achieved by using bespoke optomechanical components. 
	
	Similarly to the case of polarization-only encoding, the transition towards long-distance applications of entanglement encoded in non-polarization degrees of freedom will be challenged by increased loss and noise on the channel. Our scheme relies on quantum steering, which is more tolerant to loss and noise than Bell tests. The scheme could tolerate a larger channel loss to an extent by increasing the number of measurement settings, but longer distances will eventually require the use of a quantum repeater and loss- and noise-tolerant schemes. Our lab-scale demonstration of detection loophole-free entanglement verification shows the feasibility of using non-polarization degrees of freedom in high-demand applications. We hope it will stimulate further research and development of long-distance communication schemes that benefit from the use of different degrees of freedom of photons.
	
	The measurement alignment condition that is overcome by the use of vector vortex states can, in principle, also be relaxed by using specially-designed entanglement verification procedures based on extra and/or careful measurement direction selections~\cite{palsson12,shadbolt12,wollmann18}. However, the use of a physical encoding with total zero angular momentum, as in Eq.~(\ref{states}), offers a natural method that can be used with any conventional entanglement test. In addition, we note that the resilience of vector vortex states to misalignment and turbulence makes them compatible with event-ready verification schemes~\cite{weston18,tsujimoto20}, opening the door to new free-space quantum relay architectures for device-independent communication. Finally, the q-plate-based encoding method can be extended to qudits~\cite{milione15ol}, which, although not rotation-invariant, may enable loophole-free high-dimensional quantum steering over a fiber or a free-space channel.

	\section{Acknowledgements}
	\noindent This work was conducted by the ARC Centre of Excellence for Quantum Computation and Communication Technology under grant CE170100012. D.J.J. and F.G. acknowledge support by the  Australian Government Research Training Program (RTP).
	\vspace{1 EM}

	%	\bibliography{bibliography}

\begin{thebibliography}{67}%
		\makeatletter
		\providecommand \@ifxundefined [1]{%
			\@ifx{#1\undefined}
		}%
		\providecommand \@ifnum [1]{%
			\ifnum #1\expandafter \@firstoftwo
			\else \expandafter \@secondoftwo
			\fi
		}%
		\providecommand \@ifx [1]{%
			\ifx #1\expandafter \@firstoftwo
			\else \expandafter \@secondoftwo
			\fi
		}%
		\providecommand \natexlab [1]{#1}%
		\providecommand \enquote  [1]{``#1''}%
		\providecommand \bibnamefont  [1]{#1}%
		\providecommand \bibfnamefont [1]{#1}%
		\providecommand \citenamefont [1]{#1}%
		\providecommand \href@noop [0]{\@secondoftwo}%
		\providecommand \href [0]{\begingroup \@sanitize@url \@href}%
		\providecommand \@href[1]{\@@startlink{#1}\@@href}%
		\providecommand \@@href[1]{\endgroup#1\@@endlink}%
		\providecommand \@sanitize@url [0]{\catcode `\\12\catcode `\$12\catcode
			`\&12\catcode `\#12\catcode `\^12\catcode `\_12\catcode `\%12\relax}%
		\providecommand \@@startlink[1]{}%
		\providecommand \@@endlink[0]{}%
		\providecommand \url  [0]{\begingroup\@sanitize@url \@url }%
		\providecommand \@url [1]{\endgroup\@href {#1}{\urlprefix }}%
		\providecommand \urlprefix  [0]{URL }%
		\providecommand \Eprint [0]{\href }%
		\providecommand \doibase [0]{http://dx.doi.org/}%
		\providecommand \selectlanguage [0]{\@gobble}%
		\providecommand \bibinfo  [0]{\@secondoftwo}%
		\providecommand \bibfield  [0]{\@secondoftwo}%
		\providecommand \translation [1]{[#1]}%
		\providecommand \BibitemOpen [0]{}%
		\providecommand \bibitemStop [0]{}%
		\providecommand \bibitemNoStop [0]{.\EOS\space}%
		\providecommand \EOS [0]{\spacefactor3000\relax}%
		\providecommand \BibitemShut  [1]{\csname bibitem#1\endcsname}%
		\let\auto@bib@innerbib\@empty
		%</preamble>
		\bibitem [{\citenamefont {Flamini}\ \emph {et~al.}(2019)\citenamefont
			{Flamini}, \citenamefont {Spagnolo},\ and\ \citenamefont
			{Sciarrino}}]{rev_flamini18}%
		\BibitemOpen
		\bibfield  {author} {\bibinfo {author} {\bibfnamefont {F.}~\bibnamefont
				{Flamini}}, \bibinfo {author} {\bibfnamefont {N.}~\bibnamefont {Spagnolo}}, \
			and\ \bibinfo {author} {\bibfnamefont {F.}~\bibnamefont {Sciarrino}},\ }\href
		{http://stacks.iop.org/0034-4885/82/i=1/a=016001} {\bibfield  {journal}
			{\bibinfo  {journal} {Rep. Prog. Phys.}\ }\textbf {\bibinfo {volume} {82}},\
			\bibinfo {pages} {016001} (\bibinfo {year} {2019})}\BibitemShut {NoStop}%
		\bibitem [{\citenamefont {Slussarenko}\ \emph {et~al.}(2017)\citenamefont
			{Slussarenko}, \citenamefont {Weston}, \citenamefont {Chrzanowski},
			\citenamefont {Shalm}, \citenamefont {Verma}, \citenamefont {Nam},\ and\
			\citenamefont {Pryde}}]{slussarenko17n}%
		\BibitemOpen
		\bibfield  {author} {\bibinfo {author} {\bibfnamefont {S.}~\bibnamefont
				{Slussarenko}}, \bibinfo {author} {\bibfnamefont {M.~M.}\ \bibnamefont
				{Weston}}, \bibinfo {author} {\bibfnamefont {H.~M.}\ \bibnamefont
				{Chrzanowski}}, \bibinfo {author} {\bibfnamefont {L.~K.}\ \bibnamefont
				{Shalm}}, \bibinfo {author} {\bibfnamefont {V.~B.}\ \bibnamefont {Verma}},
			\bibinfo {author} {\bibfnamefont {S.~W.}\ \bibnamefont {Nam}}, \ and\
			\bibinfo {author} {\bibfnamefont {G.~J.}\ \bibnamefont {Pryde}},\ }\href
		{https://doi.org/10.1038/s41566-017-0011-5} {\bibfield  {journal} {\bibinfo
				{journal} {Nat. Photon.}\ }\textbf {\bibinfo {volume} {11}},\ \bibinfo
			{pages} {700} (\bibinfo {year} {2017})}\BibitemShut {NoStop}%
		\bibitem [{\citenamefont {Slussarenko}\ and\ \citenamefont
			{Pryde}(2019)}]{rev_slussarenko19}%
		\BibitemOpen
		\bibfield  {author} {\bibinfo {author} {\bibfnamefont {S.}~\bibnamefont
				{Slussarenko}}\ and\ \bibinfo {author} {\bibfnamefont {G.~J.}\ \bibnamefont
				{Pryde}},\ }\href {\doibase 10.1063/1.5115814} {\bibfield  {journal}
			{\bibinfo  {journal} {Appl. Phys. Rev.}\ }\textbf {\bibinfo {volume} {6}},\
			\bibinfo {pages} {041303} (\bibinfo {year} {2019})}\BibitemShut {NoStop}%
		\bibitem [{\citenamefont {Hensen}\ \emph {et~al.}(2015)\citenamefont {Hensen},
			\citenamefont {Bernien}, \citenamefont {Dreau}, \citenamefont {Reiserer},
			\citenamefont {Kalb}, \citenamefont {Blok}, \citenamefont {Ruitenberg},
			\citenamefont {Vermeulen}, \citenamefont {Schouten}, \citenamefont {Abellan},
			\citenamefont {Amaya}, \citenamefont {Pruneri}, \citenamefont {Mitchell},
			\citenamefont {Markham}, \citenamefont {Twitchen}, \citenamefont {Elkouss},
			\citenamefont {Wehner}, \citenamefont {Taminiau},\ and\ \citenamefont
			{Hanson}}]{hensen15}%
		\BibitemOpen
		\bibfield  {author} {\bibinfo {author} {\bibfnamefont {B.}~\bibnamefont
				{Hensen}}, \bibinfo {author} {\bibfnamefont {H.}~\bibnamefont {Bernien}},
			\bibinfo {author} {\bibfnamefont {A.~E.}\ \bibnamefont {Dreau}}, \bibinfo
			{author} {\bibfnamefont {A.}~\bibnamefont {Reiserer}}, \bibinfo {author}
			{\bibfnamefont {N.}~\bibnamefont {Kalb}}, \bibinfo {author} {\bibfnamefont
				{M.~S.}\ \bibnamefont {Blok}}, \bibinfo {author} {\bibfnamefont
				{J.}~\bibnamefont {Ruitenberg}}, \bibinfo {author} {\bibfnamefont {R.~F.~L.}\
				\bibnamefont {Vermeulen}}, \bibinfo {author} {\bibfnamefont {R.~N.}\
				\bibnamefont {Schouten}}, \bibinfo {author} {\bibfnamefont {C.}~\bibnamefont
				{Abellan}}, \bibinfo {author} {\bibfnamefont {W.}~\bibnamefont {Amaya}},
			\bibinfo {author} {\bibfnamefont {V.}~\bibnamefont {Pruneri}}, \bibinfo
			{author} {\bibfnamefont {M.~W.}\ \bibnamefont {Mitchell}}, \bibinfo {author}
			{\bibfnamefont {M.}~\bibnamefont {Markham}}, \bibinfo {author} {\bibfnamefont
				{D.~J.}\ \bibnamefont {Twitchen}}, \bibinfo {author} {\bibfnamefont
				{D.}~\bibnamefont {Elkouss}}, \bibinfo {author} {\bibfnamefont
				{S.}~\bibnamefont {Wehner}}, \bibinfo {author} {\bibfnamefont {T.~H.}\
				\bibnamefont {Taminiau}}, \ and\ \bibinfo {author} {\bibfnamefont
				{R.}~\bibnamefont {Hanson}},\ }\href {http://dx.doi.org/10.1038/nature15759}
		{\bibfield  {journal} {\bibinfo  {journal} {Nature}\ }\textbf {\bibinfo
				{volume} {526}},\ \bibinfo {pages} {682} (\bibinfo {year}
			{2015})}\BibitemShut {NoStop}%
		\bibitem [{\citenamefont {Giustina}\ \emph {et~al.}(2015)\citenamefont
			{Giustina}, \citenamefont {Versteegh}, \citenamefont {Wengerowsky},
			\citenamefont {Handsteiner}, \citenamefont {Hochrainer}, \citenamefont
			{Phelan}, \citenamefont {Steinlechner}, \citenamefont {Kofler}, \citenamefont
			{Larsson}, \citenamefont {Abell\'an}, \citenamefont {Amaya}, \citenamefont
			{Pruneri}, \citenamefont {Mitchell}, \citenamefont {Beyer}, \citenamefont
			{Gerrits}, \citenamefont {Lita}, \citenamefont {Shalm}, \citenamefont {Nam},
			\citenamefont {Scheidl}, \citenamefont {Ursin}, \citenamefont {Wittmann},\
			and\ \citenamefont {Zeilinger}}]{giustina15}%
		\BibitemOpen
		\bibfield  {author} {\bibinfo {author} {\bibfnamefont {M.}~\bibnamefont
				{Giustina}}, \bibinfo {author} {\bibfnamefont {M.~A.~M.}\ \bibnamefont
				{Versteegh}}, \bibinfo {author} {\bibfnamefont {S.}~\bibnamefont
				{Wengerowsky}}, \bibinfo {author} {\bibfnamefont {J.}~\bibnamefont
				{Handsteiner}}, \bibinfo {author} {\bibfnamefont {A.}~\bibnamefont
				{Hochrainer}}, \bibinfo {author} {\bibfnamefont {K.}~\bibnamefont {Phelan}},
			\bibinfo {author} {\bibfnamefont {F.}~\bibnamefont {Steinlechner}}, \bibinfo
			{author} {\bibfnamefont {J.}~\bibnamefont {Kofler}}, \bibinfo {author}
			{\bibfnamefont {J.-{\AA}.}\ \bibnamefont {Larsson}}, \bibinfo {author}
			{\bibfnamefont {C.}~\bibnamefont {Abell\'an}}, \bibinfo {author}
			{\bibfnamefont {W.}~\bibnamefont {Amaya}}, \bibinfo {author} {\bibfnamefont
				{V.}~\bibnamefont {Pruneri}}, \bibinfo {author} {\bibfnamefont {M.~W.}\
				\bibnamefont {Mitchell}}, \bibinfo {author} {\bibfnamefont {J.}~\bibnamefont
				{Beyer}}, \bibinfo {author} {\bibfnamefont {T.}~\bibnamefont {Gerrits}},
			\bibinfo {author} {\bibfnamefont {A.~E.}\ \bibnamefont {Lita}}, \bibinfo
			{author} {\bibfnamefont {L.~K.}\ \bibnamefont {Shalm}}, \bibinfo {author}
			{\bibfnamefont {S.~W.}\ \bibnamefont {Nam}}, \bibinfo {author} {\bibfnamefont
				{T.}~\bibnamefont {Scheidl}}, \bibinfo {author} {\bibfnamefont
				{R.}~\bibnamefont {Ursin}}, \bibinfo {author} {\bibfnamefont
				{B.}~\bibnamefont {Wittmann}}, \ and\ \bibinfo {author} {\bibfnamefont
				{A.}~\bibnamefont {Zeilinger}},\ }\href {\doibase
			10.1103/PhysRevLett.115.250401} {\bibfield  {journal} {\bibinfo  {journal}
				{Phys. Rev. Lett.}\ }\textbf {\bibinfo {volume} {115}},\ \bibinfo {pages}
			{250401} (\bibinfo {year} {2015})}\BibitemShut {NoStop}%
		\bibitem [{\citenamefont {Shalm}\ \emph {et~al.}(2015)\citenamefont {Shalm},
			\citenamefont {Meyer-Scott}, \citenamefont {Christensen}, \citenamefont
			{Bierhorst}, \citenamefont {Wayne}, \citenamefont {Stevens}, \citenamefont
			{Gerrits}, \citenamefont {Glancy}, \citenamefont {Hamel}, \citenamefont
			{Allman}, \citenamefont {Coakley}, \citenamefont {Dyer}, \citenamefont
			{Hodge}, \citenamefont {Lita}, \citenamefont {Verma}, \citenamefont
			{Lambrocco}, \citenamefont {Tortorici}, \citenamefont {Migdall},
			\citenamefont {Zhang}, \citenamefont {Kumor}, \citenamefont {Farr},
			\citenamefont {Marsili}, \citenamefont {Shaw}, \citenamefont {Stern},
			\citenamefont {Abell\'an}, \citenamefont {Amaya}, \citenamefont {Pruneri},
			\citenamefont {Jennewein}, \citenamefont {Mitchell}, \citenamefont {Kwiat},
			\citenamefont {Bienfang}, \citenamefont {Mirin}, \citenamefont {Knill},\ and\
			\citenamefont {Nam}}]{shalm15}%
		\BibitemOpen
		\bibfield  {author} {\bibinfo {author} {\bibfnamefont {L.~K.}\ \bibnamefont
				{Shalm}}, \bibinfo {author} {\bibfnamefont {E.}~\bibnamefont {Meyer-Scott}},
			\bibinfo {author} {\bibfnamefont {B.~G.}\ \bibnamefont {Christensen}},
			\bibinfo {author} {\bibfnamefont {P.}~\bibnamefont {Bierhorst}}, \bibinfo
			{author} {\bibfnamefont {M.~A.}\ \bibnamefont {Wayne}}, \bibinfo {author}
			{\bibfnamefont {M.~J.}\ \bibnamefont {Stevens}}, \bibinfo {author}
			{\bibfnamefont {T.}~\bibnamefont {Gerrits}}, \bibinfo {author} {\bibfnamefont
				{S.}~\bibnamefont {Glancy}}, \bibinfo {author} {\bibfnamefont {D.~R.}\
				\bibnamefont {Hamel}}, \bibinfo {author} {\bibfnamefont {M.~S.}\ \bibnamefont
				{Allman}}, \bibinfo {author} {\bibfnamefont {K.~J.}\ \bibnamefont {Coakley}},
			\bibinfo {author} {\bibfnamefont {S.~D.}\ \bibnamefont {Dyer}}, \bibinfo
			{author} {\bibfnamefont {C.}~\bibnamefont {Hodge}}, \bibinfo {author}
			{\bibfnamefont {A.~E.}\ \bibnamefont {Lita}}, \bibinfo {author}
			{\bibfnamefont {V.~B.}\ \bibnamefont {Verma}}, \bibinfo {author}
			{\bibfnamefont {C.}~\bibnamefont {Lambrocco}}, \bibinfo {author}
			{\bibfnamefont {E.}~\bibnamefont {Tortorici}}, \bibinfo {author}
			{\bibfnamefont {A.~L.}\ \bibnamefont {Migdall}}, \bibinfo {author}
			{\bibfnamefont {Y.}~\bibnamefont {Zhang}}, \bibinfo {author} {\bibfnamefont
				{D.~R.}\ \bibnamefont {Kumor}}, \bibinfo {author} {\bibfnamefont {W.~H.}\
				\bibnamefont {Farr}}, \bibinfo {author} {\bibfnamefont {F.}~\bibnamefont
				{Marsili}}, \bibinfo {author} {\bibfnamefont {M.~D.}\ \bibnamefont {Shaw}},
			\bibinfo {author} {\bibfnamefont {J.~A.}\ \bibnamefont {Stern}}, \bibinfo
			{author} {\bibfnamefont {C.}~\bibnamefont {Abell\'an}}, \bibinfo {author}
			{\bibfnamefont {W.}~\bibnamefont {Amaya}}, \bibinfo {author} {\bibfnamefont
				{V.}~\bibnamefont {Pruneri}}, \bibinfo {author} {\bibfnamefont
				{T.}~\bibnamefont {Jennewein}}, \bibinfo {author} {\bibfnamefont {M.~W.}\
				\bibnamefont {Mitchell}}, \bibinfo {author} {\bibfnamefont {P.~G.}\
				\bibnamefont {Kwiat}}, \bibinfo {author} {\bibfnamefont {J.~C.}\ \bibnamefont
				{Bienfang}}, \bibinfo {author} {\bibfnamefont {R.~P.}\ \bibnamefont {Mirin}},
			\bibinfo {author} {\bibfnamefont {E.}~\bibnamefont {Knill}}, \ and\ \bibinfo
			{author} {\bibfnamefont {S.~W.}\ \bibnamefont {Nam}},\ }\href {\doibase
			10.1103/PhysRevLett.115.250402} {\bibfield  {journal} {\bibinfo  {journal}
				{Phys. Rev. Lett.}\ }\textbf {\bibinfo {volume} {115}},\ \bibinfo {pages}
			{250402} (\bibinfo {year} {2015})}\BibitemShut {NoStop}%
		\bibitem [{\citenamefont {Collaboration}(2018)}]{bbt18}%
		\BibitemOpen
		\bibfield  {author} {\bibinfo {author} {\bibfnamefont {T.~B.~B.}\
				\bibnamefont {Collaboration}},\ }\href
		{https://doi.org/10.1038/s41586-018-0085-3} {\bibfield  {journal} {\bibinfo
				{journal} {Nature}\ }\textbf {\bibinfo {volume} {557}},\ \bibinfo {pages}
			{212} (\bibinfo {year} {2018})}\BibitemShut {NoStop}%
		\bibitem [{\citenamefont {Rauch}\ \emph {et~al.}(2018)\citenamefont {Rauch},
			\citenamefont {Handsteiner}, \citenamefont {Hochrainer}, \citenamefont
			{Gallicchio}, \citenamefont {Friedman}, \citenamefont {Leung}, \citenamefont
			{Liu}, \citenamefont {Bulla}, \citenamefont {Ecker}, \citenamefont
			{Steinlechner}, \citenamefont {Ursin}, \citenamefont {Hu}, \citenamefont
			{Leon}, \citenamefont {Benn}, \citenamefont {Ghedina}, \citenamefont
			{Cecconi}, \citenamefont {Guth}, \citenamefont {Kaiser}, \citenamefont
			{Scheidl},\ and\ \citenamefont {Zeilinger}}]{rauch18}%
		\BibitemOpen
		\bibfield  {author} {\bibinfo {author} {\bibfnamefont {D.}~\bibnamefont
				{Rauch}}, \bibinfo {author} {\bibfnamefont {J.}~\bibnamefont {Handsteiner}},
			\bibinfo {author} {\bibfnamefont {A.}~\bibnamefont {Hochrainer}}, \bibinfo
			{author} {\bibfnamefont {J.}~\bibnamefont {Gallicchio}}, \bibinfo {author}
			{\bibfnamefont {A.~S.}\ \bibnamefont {Friedman}}, \bibinfo {author}
			{\bibfnamefont {C.}~\bibnamefont {Leung}}, \bibinfo {author} {\bibfnamefont
				{B.}~\bibnamefont {Liu}}, \bibinfo {author} {\bibfnamefont {L.}~\bibnamefont
				{Bulla}}, \bibinfo {author} {\bibfnamefont {S.}~\bibnamefont {Ecker}},
			\bibinfo {author} {\bibfnamefont {F.}~\bibnamefont {Steinlechner}}, \bibinfo
			{author} {\bibfnamefont {R.}~\bibnamefont {Ursin}}, \bibinfo {author}
			{\bibfnamefont {B.}~\bibnamefont {Hu}}, \bibinfo {author} {\bibfnamefont
				{D.}~\bibnamefont {Leon}}, \bibinfo {author} {\bibfnamefont {C.}~\bibnamefont
				{Benn}}, \bibinfo {author} {\bibfnamefont {A.}~\bibnamefont {Ghedina}},
			\bibinfo {author} {\bibfnamefont {M.}~\bibnamefont {Cecconi}}, \bibinfo
			{author} {\bibfnamefont {A.~H.}\ \bibnamefont {Guth}}, \bibinfo {author}
			{\bibfnamefont {D.~I.}\ \bibnamefont {Kaiser}}, \bibinfo {author}
			{\bibfnamefont {T.}~\bibnamefont {Scheidl}}, \ and\ \bibinfo {author}
			{\bibfnamefont {A.}~\bibnamefont {Zeilinger}},\ }\href {\doibase
			10.1103/PhysRevLett.121.080403} {\bibfield  {journal} {\bibinfo  {journal}
				{Phys. Rev. Lett.}\ }\textbf {\bibinfo {volume} {121}},\ \bibinfo {pages}
			{080403} (\bibinfo {year} {2018})}\BibitemShut {NoStop}%
		\bibitem [{\citenamefont {Wittmann}\ \emph {et~al.}(2012)\citenamefont
			{Wittmann}, \citenamefont {Ramelow}, \citenamefont {Steinlechner},
			\citenamefont {Langford}, \citenamefont {Brunner}, \citenamefont {Wiseman},
			\citenamefont {Ursin},\ and\ \citenamefont {Zeilinger}}]{wittmann12}%
		\BibitemOpen
		\bibfield  {author} {\bibinfo {author} {\bibfnamefont {B.}~\bibnamefont
				{Wittmann}}, \bibinfo {author} {\bibfnamefont {S.}~\bibnamefont {Ramelow}},
			\bibinfo {author} {\bibfnamefont {F.}~\bibnamefont {Steinlechner}}, \bibinfo
			{author} {\bibfnamefont {N.~K.}\ \bibnamefont {Langford}}, \bibinfo {author}
			{\bibfnamefont {N.}~\bibnamefont {Brunner}}, \bibinfo {author} {\bibfnamefont
				{H.~M.}\ \bibnamefont {Wiseman}}, \bibinfo {author} {\bibfnamefont
				{R.}~\bibnamefont {Ursin}}, \ and\ \bibinfo {author} {\bibfnamefont
				{A.}~\bibnamefont {Zeilinger}},\ }\href
		{http://stacks.iop.org/1367-2630/14/i=5/a=053030} {\bibfield  {journal}
			{\bibinfo  {journal} {New J. Phys.}\ }\textbf {\bibinfo {volume} {14}},\
			\bibinfo {pages} {053030} (\bibinfo {year} {2012})}\BibitemShut {NoStop}%
		\bibitem [{\citenamefont {Ac\'{i}n}\ \emph {et~al.}(2006)\citenamefont
			{Ac\'{i}n}, \citenamefont {Gisin},\ and\ \citenamefont {Masanes}}]{acin06}%
		\BibitemOpen
		\bibfield  {author} {\bibinfo {author} {\bibfnamefont {A.}~\bibnamefont
				{Ac\'{i}n}}, \bibinfo {author} {\bibfnamefont {N.}~\bibnamefont {Gisin}}, \
			and\ \bibinfo {author} {\bibfnamefont {L.}~\bibnamefont {Masanes}},\ }\href
		{\doibase 10.1103/PhysRevLett.97.120405} {\bibfield  {journal} {\bibinfo
				{journal} {Phys. Rev. Lett.}\ }\textbf {\bibinfo {volume} {97}},\ \bibinfo
			{pages} {120405} (\bibinfo {year} {2006})}\BibitemShut {NoStop}%
		\bibitem [{\citenamefont {Branciard}\ \emph {et~al.}(2012)\citenamefont
			{Branciard}, \citenamefont {Cavalcanti}, \citenamefont {Walborn},
			\citenamefont {Scarani},\ and\ \citenamefont {Wiseman}}]{branciard12}%
		\BibitemOpen
		\bibfield  {author} {\bibinfo {author} {\bibfnamefont {C.}~\bibnamefont
				{Branciard}}, \bibinfo {author} {\bibfnamefont {E.~G.}\ \bibnamefont
				{Cavalcanti}}, \bibinfo {author} {\bibfnamefont {S.~P.}\ \bibnamefont
				{Walborn}}, \bibinfo {author} {\bibfnamefont {V.}~\bibnamefont {Scarani}}, \
			and\ \bibinfo {author} {\bibfnamefont {H.~M.}\ \bibnamefont {Wiseman}},\
		}\href {\doibase 10.1103/PhysRevA.85.010301} {\bibfield  {journal} {\bibinfo
				{journal} {Phys. Rev. A}\ }\textbf {\bibinfo {volume} {85}},\ \bibinfo
			{pages} {010301} (\bibinfo {year} {2012})}\BibitemShut {NoStop}%
		\bibitem [{\citenamefont {Law}\ \emph {et~al.}(2014)\citenamefont {Law},
			\citenamefont {Thinh}, \citenamefont {Bancal},\ and\ \citenamefont
			{Scarani}}]{law14}%
		\BibitemOpen
		\bibfield  {author} {\bibinfo {author} {\bibfnamefont {Y.~Z.}\ \bibnamefont
				{Law}}, \bibinfo {author} {\bibfnamefont {L.~P.}\ \bibnamefont {Thinh}},
			\bibinfo {author} {\bibfnamefont {J.-D.}\ \bibnamefont {Bancal}}, \ and\
			\bibinfo {author} {\bibfnamefont {V.}~\bibnamefont {Scarani}},\ }\href
		{\doibase 10.1088/1751-8113/47/42/424028} {\bibfield  {journal} {\bibinfo
				{journal} {J. Phys. A: Math. Theor.}\ }\textbf {\bibinfo {volume} {47}},\
			\bibinfo {pages} {424028} (\bibinfo {year} {2014})}\BibitemShut {NoStop}%
		\bibitem [{\citenamefont {Bierhorst}\ \emph {et~al.}(2018)\citenamefont
			{Bierhorst}, \citenamefont {Knill}, \citenamefont {Glancy}, \citenamefont
			{Zhang}, \citenamefont {Mink}, \citenamefont {Jordan}, \citenamefont
			{Rommal}, \citenamefont {Liu}, \citenamefont {Christensen}, \citenamefont
			{Nam}, \citenamefont {Stevens},\ and\ \citenamefont {Shalm}}]{bierhorst18}%
		\BibitemOpen
		\bibfield  {author} {\bibinfo {author} {\bibfnamefont {P.}~\bibnamefont
				{Bierhorst}}, \bibinfo {author} {\bibfnamefont {E.}~\bibnamefont {Knill}},
			\bibinfo {author} {\bibfnamefont {S.}~\bibnamefont {Glancy}}, \bibinfo
			{author} {\bibfnamefont {Y.}~\bibnamefont {Zhang}}, \bibinfo {author}
			{\bibfnamefont {A.}~\bibnamefont {Mink}}, \bibinfo {author} {\bibfnamefont
				{S.}~\bibnamefont {Jordan}}, \bibinfo {author} {\bibfnamefont
				{A.}~\bibnamefont {Rommal}}, \bibinfo {author} {\bibfnamefont {Y.-K.}\
				\bibnamefont {Liu}}, \bibinfo {author} {\bibfnamefont {B.}~\bibnamefont
				{Christensen}}, \bibinfo {author} {\bibfnamefont {S.~W.}\ \bibnamefont
				{Nam}}, \bibinfo {author} {\bibfnamefont {M.~J.}\ \bibnamefont {Stevens}}, \
			and\ \bibinfo {author} {\bibfnamefont {L.~K.}\ \bibnamefont {Shalm}},\ }\href
		{https://doi.org/10.1038/s41586-018-0019-0} {\bibfield  {journal} {\bibinfo
				{journal} {Nature}\ }\textbf {\bibinfo {volume} {556}},\ \bibinfo {pages}
			{223} (\bibinfo {year} {2018})}\BibitemShut {NoStop}%
		\bibitem [{\citenamefont {Larsson}(2014)}]{larsson14}%
		\BibitemOpen
		\bibfield  {author} {\bibinfo {author} {\bibfnamefont {J.-A.}\ \bibnamefont
				{Larsson}},\ }\href {http://stacks.iop.org/1751-8121/47/i=42/a=424003}
		{\bibfield  {journal} {\bibinfo  {journal} {J. Phys. A}\ }\textbf {\bibinfo
				{volume} {47}},\ \bibinfo {pages} {424003} (\bibinfo {year}
			{2014})}\BibitemShut {NoStop}%
		\bibitem [{\citenamefont {Bennet}\ \emph {et~al.}(2012)\citenamefont {Bennet},
			\citenamefont {Evans}, \citenamefont {Saunders}, \citenamefont {Branciard},
			\citenamefont {Cavalcanti}, \citenamefont {Wiseman},\ and\ \citenamefont
			{Pryde}}]{bennet12}%
		\BibitemOpen
		\bibfield  {author} {\bibinfo {author} {\bibfnamefont {A.~J.}\ \bibnamefont
				{Bennet}}, \bibinfo {author} {\bibfnamefont {D.~A.}\ \bibnamefont {Evans}},
			\bibinfo {author} {\bibfnamefont {D.~J.}\ \bibnamefont {Saunders}}, \bibinfo
			{author} {\bibfnamefont {C.}~\bibnamefont {Branciard}}, \bibinfo {author}
			{\bibfnamefont {E.~G.}\ \bibnamefont {Cavalcanti}}, \bibinfo {author}
			{\bibfnamefont {H.~M.}\ \bibnamefont {Wiseman}}, \ and\ \bibinfo {author}
			{\bibfnamefont {G.~J.}\ \bibnamefont {Pryde}},\ }\href {\doibase
			10.1103/PhysRevX.2.031003} {\bibfield  {journal} {\bibinfo  {journal} {Phys.
					Rev. X}\ }\textbf {\bibinfo {volume} {2}},\ \bibinfo {pages} {031003}
			(\bibinfo {year} {2012})}\BibitemShut {NoStop}%
		\bibitem [{\citenamefont {Erhard}\ \emph
			{et~al.}(2018{\natexlab{a}})\citenamefont {Erhard}, \citenamefont {Fickler},
			\citenamefont {Krenn},\ and\ \citenamefont {Zeilinger}}]{rev_erhard18}%
		\BibitemOpen
		\bibfield  {author} {\bibinfo {author} {\bibfnamefont {M.}~\bibnamefont
				{Erhard}}, \bibinfo {author} {\bibfnamefont {R.}~\bibnamefont {Fickler}},
			\bibinfo {author} {\bibfnamefont {M.}~\bibnamefont {Krenn}}, \ and\ \bibinfo
			{author} {\bibfnamefont {A.}~\bibnamefont {Zeilinger}},\ }\href
		{https://doi.org/10.1038/lsa.2017.146} {\bibfield  {journal} {\bibinfo
				{journal} {Light Sci. Appl.}\ }\textbf {\bibinfo {volume} {7}},\ \bibinfo
			{pages} {17146} (\bibinfo {year} {2018}{\natexlab{a}})}\BibitemShut {NoStop}%
		\bibitem [{\citenamefont {Ansari}\ \emph {et~al.}(2018)\citenamefont {Ansari},
			\citenamefont {Roccia}, \citenamefont {Santandrea}, \citenamefont {Doostdar},
			\citenamefont {Eigner}, \citenamefont {Padberg}, \citenamefont {Gianani},
			\citenamefont {Sbroscia}, \citenamefont {Donohue}, \citenamefont {Mancino},
			\citenamefont {Barbieri},\ and\ \citenamefont {Silberhorn}}]{ansari18}%
		\BibitemOpen
		\bibfield  {author} {\bibinfo {author} {\bibfnamefont {V.}~\bibnamefont
				{Ansari}}, \bibinfo {author} {\bibfnamefont {E.}~\bibnamefont {Roccia}},
			\bibinfo {author} {\bibfnamefont {M.}~\bibnamefont {Santandrea}}, \bibinfo
			{author} {\bibfnamefont {M.}~\bibnamefont {Doostdar}}, \bibinfo {author}
			{\bibfnamefont {C.}~\bibnamefont {Eigner}}, \bibinfo {author} {\bibfnamefont
				{L.}~\bibnamefont {Padberg}}, \bibinfo {author} {\bibfnamefont
				{I.}~\bibnamefont {Gianani}}, \bibinfo {author} {\bibfnamefont
				{M.}~\bibnamefont {Sbroscia}}, \bibinfo {author} {\bibfnamefont {J.~M.}\
				\bibnamefont {Donohue}}, \bibinfo {author} {\bibfnamefont {L.}~\bibnamefont
				{Mancino}}, \bibinfo {author} {\bibfnamefont {M.}~\bibnamefont {Barbieri}}, \
			and\ \bibinfo {author} {\bibfnamefont {C.}~\bibnamefont {Silberhorn}},\
		}\href {\doibase 10.1364/OE.26.002764} {\bibfield  {journal} {\bibinfo
				{journal} {Opt. Express}\ }\textbf {\bibinfo {volume} {26}},\ \bibinfo
			{pages} {2764} (\bibinfo {year} {2018})}\BibitemShut {NoStop}%
		\bibitem [{\citenamefont {Cai}\ \emph {et~al.}(2017)\citenamefont {Cai},
			\citenamefont {Roslund}, \citenamefont {Ferrini}, \citenamefont {Arzani},
			\citenamefont {Xu}, \citenamefont {Fabre},\ and\ \citenamefont
			{Treps}}]{cai17}%
		\BibitemOpen
		\bibfield  {author} {\bibinfo {author} {\bibfnamefont {Y.}~\bibnamefont
				{Cai}}, \bibinfo {author} {\bibfnamefont {J.}~\bibnamefont {Roslund}},
			\bibinfo {author} {\bibfnamefont {G.}~\bibnamefont {Ferrini}}, \bibinfo
			{author} {\bibfnamefont {F.}~\bibnamefont {Arzani}}, \bibinfo {author}
			{\bibfnamefont {X.}~\bibnamefont {Xu}}, \bibinfo {author} {\bibfnamefont
				{C.}~\bibnamefont {Fabre}}, \ and\ \bibinfo {author} {\bibfnamefont
				{N.}~\bibnamefont {Treps}},\ }\href {https://doi.org/10.1038/ncomms15645}
		{\bibfield  {journal} {\bibinfo  {journal} {Nat. Commun.}\ }\textbf {\bibinfo
				{volume} {8}},\ \bibinfo {pages} {15645} (\bibinfo {year}
			{2017})}\BibitemShut {NoStop}%
		\bibitem [{\citenamefont {Korzh}\ \emph {et~al.}(2015)\citenamefont {Korzh},
			\citenamefont {Lim}, \citenamefont {Houlmann}, \citenamefont {Gisin},
			\citenamefont {Li}, \citenamefont {Nolan}, \citenamefont {Sanguinetti},
			\citenamefont {Thew},\ and\ \citenamefont {Zbinden}}]{korzh15}%
		\BibitemOpen
		\bibfield  {author} {\bibinfo {author} {\bibfnamefont {B.}~\bibnamefont
				{Korzh}}, \bibinfo {author} {\bibfnamefont {C.~C.~W.}\ \bibnamefont {Lim}},
			\bibinfo {author} {\bibfnamefont {R.}~\bibnamefont {Houlmann}}, \bibinfo
			{author} {\bibfnamefont {N.}~\bibnamefont {Gisin}}, \bibinfo {author}
			{\bibfnamefont {M.~J.}\ \bibnamefont {Li}}, \bibinfo {author} {\bibfnamefont
				{D.}~\bibnamefont {Nolan}}, \bibinfo {author} {\bibfnamefont
				{B.}~\bibnamefont {Sanguinetti}}, \bibinfo {author} {\bibfnamefont
				{R.}~\bibnamefont {Thew}}, \ and\ \bibinfo {author} {\bibfnamefont
				{H.}~\bibnamefont {Zbinden}},\ }\href
		{http://dx.doi.org/10.1038/nphoton.2014.327} {\bibfield  {journal} {\bibinfo
				{journal} {Nat. Photon.}\ }\textbf {\bibinfo {volume} {9}},\ \bibinfo {pages}
			{163} (\bibinfo {year} {2015})}\BibitemShut {NoStop}%
		\bibitem [{\citenamefont {Ecker}\ \emph {et~al.}(2019)\citenamefont {Ecker},
			\citenamefont {Bouchard}, \citenamefont {Bulla}, \citenamefont {Brandt},
			\citenamefont {Kohout}, \citenamefont {Steinlechner}, \citenamefont
			{Fickler}, \citenamefont {Malik}, \citenamefont {Guryanova}, \citenamefont
			{Ursin},\ and\ \citenamefont {Huber}}]{ecker19}%
		\BibitemOpen
		\bibfield  {author} {\bibinfo {author} {\bibfnamefont {S.}~\bibnamefont
				{Ecker}}, \bibinfo {author} {\bibfnamefont {F.}~\bibnamefont {Bouchard}},
			\bibinfo {author} {\bibfnamefont {L.}~\bibnamefont {Bulla}}, \bibinfo
			{author} {\bibfnamefont {F.}~\bibnamefont {Brandt}}, \bibinfo {author}
			{\bibfnamefont {O.}~\bibnamefont {Kohout}}, \bibinfo {author} {\bibfnamefont
				{F.}~\bibnamefont {Steinlechner}}, \bibinfo {author} {\bibfnamefont
				{R.}~\bibnamefont {Fickler}}, \bibinfo {author} {\bibfnamefont
				{M.}~\bibnamefont {Malik}}, \bibinfo {author} {\bibfnamefont
				{Y.}~\bibnamefont {Guryanova}}, \bibinfo {author} {\bibfnamefont
				{R.}~\bibnamefont {Ursin}}, \ and\ \bibinfo {author} {\bibfnamefont
				{M.}~\bibnamefont {Huber}},\ }\href {\doibase 10.1103/PhysRevX.9.041042}
		{\bibfield  {journal} {\bibinfo  {journal} {Phys. Rev. X}\ }\textbf {\bibinfo
				{volume} {9}},\ \bibinfo {pages} {041042} (\bibinfo {year}
			{2019})}\BibitemShut {NoStop}%
		\bibitem [{\citenamefont {Zhu}\ \emph {et~al.}(2021)\citenamefont {Zhu},
			\citenamefont {Tyler}, \citenamefont {Valencia}, \citenamefont {Malik},\ and\
			\citenamefont {Leach}}]{zhu21}%
		\BibitemOpen
		\bibfield  {author} {\bibinfo {author} {\bibfnamefont {F.}~\bibnamefont
				{Zhu}}, \bibinfo {author} {\bibfnamefont {M.}~\bibnamefont {Tyler}}, \bibinfo
			{author} {\bibfnamefont {N.~H.}\ \bibnamefont {Valencia}}, \bibinfo {author}
			{\bibfnamefont {M.}~\bibnamefont {Malik}}, \ and\ \bibinfo {author}
			{\bibfnamefont {J.}~\bibnamefont {Leach}},\ }\href {\doibase
			10.1116/5.0033889} {\bibfield  {journal} {\bibinfo  {journal} {AVS Quantum
					Science}\ }\textbf {\bibinfo {volume} {3}},\ \bibinfo {pages} {011401}
			(\bibinfo {year} {2021})}\BibitemShut {NoStop}%
		\bibitem [{\citenamefont {Aolita}\ and\ \citenamefont
			{Walborn}(2007)}]{aolita07}%
		\BibitemOpen
		\bibfield  {author} {\bibinfo {author} {\bibfnamefont {L.}~\bibnamefont
				{Aolita}}\ and\ \bibinfo {author} {\bibfnamefont {S.~P.}\ \bibnamefont
				{Walborn}},\ }\href {http://dx.doi.org/10.1103/PhysRevLett.98.100501}
		{\bibfield  {journal} {\bibinfo  {journal} {Phys. Rev. Lett.}\ }\textbf
			{\bibinfo {volume} {98}},\ \bibinfo {pages} {100501} (\bibinfo {year}
			{2007})}\BibitemShut {NoStop}%
		\bibitem [{\citenamefont {D'Ambrosio}\ \emph {et~al.}(2012)\citenamefont
			{D'Ambrosio}, \citenamefont {Nagali}, \citenamefont {Walborn}, \citenamefont
			{Aolita}, \citenamefont {Slussarenko}, \citenamefont {Marrucci},\ and\
			\citenamefont {Sciarrino}}]{dambrosio12natcomm}%
		\BibitemOpen
		\bibfield  {author} {\bibinfo {author} {\bibfnamefont {V.}~\bibnamefont
				{D'Ambrosio}}, \bibinfo {author} {\bibfnamefont {E.}~\bibnamefont {Nagali}},
			\bibinfo {author} {\bibfnamefont {S.~P.}\ \bibnamefont {Walborn}}, \bibinfo
			{author} {\bibfnamefont {L.}~\bibnamefont {Aolita}}, \bibinfo {author}
			{\bibfnamefont {S.}~\bibnamefont {Slussarenko}}, \bibinfo {author}
			{\bibfnamefont {L.}~\bibnamefont {Marrucci}}, \ and\ \bibinfo {author}
			{\bibfnamefont {F.}~\bibnamefont {Sciarrino}},\ }\href {\doibase
			10.1038/ncomms1951} {\bibfield  {journal} {\bibinfo  {journal} {Nat.
					Commun.}\ }\textbf {\bibinfo {volume} {3}},\ \bibinfo {pages} {961} (\bibinfo
			{year} {2012})}\BibitemShut {NoStop}%
		\bibitem [{\citenamefont {Vallone}\ \emph {et~al.}(2014)\citenamefont
			{Vallone}, \citenamefont {D'Ambrosio}, \citenamefont {Sponselli},
			\citenamefont {Slussarenko}, \citenamefont {Marrucci}, \citenamefont
			{Sciarrino},\ and\ \citenamefont {Villoresi}}]{vallone14}%
		\BibitemOpen
		\bibfield  {author} {\bibinfo {author} {\bibfnamefont {G.}~\bibnamefont
				{Vallone}}, \bibinfo {author} {\bibfnamefont {V.}~\bibnamefont {D'Ambrosio}},
			\bibinfo {author} {\bibfnamefont {A.}~\bibnamefont {Sponselli}}, \bibinfo
			{author} {\bibfnamefont {S.}~\bibnamefont {Slussarenko}}, \bibinfo {author}
			{\bibfnamefont {L.}~\bibnamefont {Marrucci}}, \bibinfo {author}
			{\bibfnamefont {F.}~\bibnamefont {Sciarrino}}, \ and\ \bibinfo {author}
			{\bibfnamefont {P.}~\bibnamefont {Villoresi}},\ }\href {\doibase
			10.1103/PhysRevLett.113.060503} {\bibfield  {journal} {\bibinfo  {journal}
				{Phys. Rev. Lett.}\ }\textbf {\bibinfo {volume} {113}},\ \bibinfo {pages}
			{060503} (\bibinfo {year} {2014})}\BibitemShut {NoStop}%
		\bibitem [{\citenamefont {Ndagano}\ \emph {et~al.}(2017)\citenamefont
			{Ndagano}, \citenamefont {Perez-Garcia}, \citenamefont {Roux}, \citenamefont
			{McLaren}, \citenamefont {Rosales-Guzman}, \citenamefont {Zhang},
			\citenamefont {Mouane}, \citenamefont {Hernandez-Aranda}, \citenamefont
			{Konrad},\ and\ \citenamefont {Forbes}}]{ndagano17}%
		\BibitemOpen
		\bibfield  {author} {\bibinfo {author} {\bibfnamefont {B.}~\bibnamefont
				{Ndagano}}, \bibinfo {author} {\bibfnamefont {B.}~\bibnamefont
				{Perez-Garcia}}, \bibinfo {author} {\bibfnamefont {F.~S.}\ \bibnamefont
				{Roux}}, \bibinfo {author} {\bibfnamefont {M.}~\bibnamefont {McLaren}},
			\bibinfo {author} {\bibfnamefont {C.}~\bibnamefont {Rosales-Guzman}},
			\bibinfo {author} {\bibfnamefont {Y.}~\bibnamefont {Zhang}}, \bibinfo
			{author} {\bibfnamefont {O.}~\bibnamefont {Mouane}}, \bibinfo {author}
			{\bibfnamefont {R.~I.}\ \bibnamefont {Hernandez-Aranda}}, \bibinfo {author}
			{\bibfnamefont {T.}~\bibnamefont {Konrad}}, \ and\ \bibinfo {author}
			{\bibfnamefont {A.}~\bibnamefont {Forbes}},\ }\href {\doibase
			10.1038/nphys4003} {\bibfield  {journal} {\bibinfo  {journal} {Nat. Phys.}\
			}\textbf {\bibinfo {volume} {13}},\ \bibinfo {pages} {397} (\bibinfo {year}
			{2017})}\BibitemShut {NoStop}%
		\bibitem [{\citenamefont {Wiseman}\ \emph {et~al.}(2007)\citenamefont
			{Wiseman}, \citenamefont {Jones},\ and\ \citenamefont {Doherty}}]{wiseman07}%
		\BibitemOpen
		\bibfield  {author} {\bibinfo {author} {\bibfnamefont {H.~M.}\ \bibnamefont
				{Wiseman}}, \bibinfo {author} {\bibfnamefont {S.~J.}\ \bibnamefont {Jones}},
			\ and\ \bibinfo {author} {\bibfnamefont {A.~C.}\ \bibnamefont {Doherty}},\
		}\href {\doibase 10.1103/PhysRevLett.98.140402} {\bibfield  {journal}
			{\bibinfo  {journal} {Phys. Rev. Lett.}\ }\textbf {\bibinfo {volume} {98}},\
			\bibinfo {pages} {140402} (\bibinfo {year} {2007})}\BibitemShut {NoStop}%
		\bibitem [{\citenamefont {Cavalcanti}\ and\ \citenamefont
			{Skrzypczyk}(2016)}]{rev_cavalcanti16}%
		\BibitemOpen
		\bibfield  {author} {\bibinfo {author} {\bibfnamefont {D.}~\bibnamefont
				{Cavalcanti}}\ and\ \bibinfo {author} {\bibfnamefont {P.}~\bibnamefont
				{Skrzypczyk}},\ }\href {\doibase 10.1088/1361-6633/80/2/024001} {\bibfield
			{journal} {\bibinfo  {journal} {Rep. Prog. Phys.}\ }\textbf {\bibinfo
				{volume} {80}},\ \bibinfo {pages} {024001} (\bibinfo {year}
			{2016})}\BibitemShut {NoStop}%
		\bibitem [{\citenamefont {Uola}\ \emph {et~al.}(2020)\citenamefont {Uola},
			\citenamefont {Costa}, \citenamefont {Nguyen},\ and\ \citenamefont
			{G\"uhne}}]{rev_uola20}%
		\BibitemOpen
		\bibfield  {author} {\bibinfo {author} {\bibfnamefont {R.}~\bibnamefont
				{Uola}}, \bibinfo {author} {\bibfnamefont {A.~C.~S.}\ \bibnamefont {Costa}},
			\bibinfo {author} {\bibfnamefont {H.~C.}\ \bibnamefont {Nguyen}}, \ and\
			\bibinfo {author} {\bibfnamefont {O.}~\bibnamefont {G\"uhne}},\ }\href
		{\doibase 10.1103/RevModPhys.92.015001} {\bibfield  {journal} {\bibinfo
				{journal} {Rev. Mod. Phys.}\ }\textbf {\bibinfo {volume} {92}},\ \bibinfo
			{pages} {015001} (\bibinfo {year} {2020})}\BibitemShut {NoStop}%
		\bibitem [{\citenamefont {Vedovato}\ \emph {et~al.}(2018)\citenamefont
			{Vedovato}, \citenamefont {Agnesi}, \citenamefont {Tomasin}, \citenamefont
			{Avesani}, \citenamefont {Larsson}, \citenamefont {Vallone},\ and\
			\citenamefont {Villoresi}}]{vedovato18}%
		\BibitemOpen
		\bibfield  {author} {\bibinfo {author} {\bibfnamefont {F.}~\bibnamefont
				{Vedovato}}, \bibinfo {author} {\bibfnamefont {C.}~\bibnamefont {Agnesi}},
			\bibinfo {author} {\bibfnamefont {M.}~\bibnamefont {Tomasin}}, \bibinfo
			{author} {\bibfnamefont {M.}~\bibnamefont {Avesani}}, \bibinfo {author}
			{\bibfnamefont {J.-A.}\ \bibnamefont {Larsson}}, \bibinfo {author}
			{\bibfnamefont {G.}~\bibnamefont {Vallone}}, \ and\ \bibinfo {author}
			{\bibfnamefont {P.}~\bibnamefont {Villoresi}},\ }\href {\doibase
			10.1103/PhysRevLett.121.190401} {\bibfield  {journal} {\bibinfo  {journal}
				{Phys. Rev. Lett.}\ }\textbf {\bibinfo {volume} {121}},\ \bibinfo {pages}
			{190401} (\bibinfo {year} {2018})}\BibitemShut {NoStop}%
		\bibitem [{\citenamefont {Herrera~Valencia}\ \emph {et~al.}(2020)\citenamefont
			{Herrera~Valencia}, \citenamefont {Srivastav}, \citenamefont {Pivoluska},
			\citenamefont {Huber}, \citenamefont {Friis}, \citenamefont {McCutcheon},\
			and\ \citenamefont {Malik}}]{valencia20}%
		\BibitemOpen
		\bibfield  {author} {\bibinfo {author} {\bibfnamefont {N.}~\bibnamefont
				{Herrera~Valencia}}, \bibinfo {author} {\bibfnamefont {V.}~\bibnamefont
				{Srivastav}}, \bibinfo {author} {\bibfnamefont {M.}~\bibnamefont
				{Pivoluska}}, \bibinfo {author} {\bibfnamefont {M.}~\bibnamefont {Huber}},
			\bibinfo {author} {\bibfnamefont {N.}~\bibnamefont {Friis}}, \bibinfo
			{author} {\bibfnamefont {W.}~\bibnamefont {McCutcheon}}, \ and\ \bibinfo
			{author} {\bibfnamefont {M.}~\bibnamefont {Malik}},\ }\href {\doibase
			10.22331/q-2020-12-24-376} {\bibfield  {journal} {\bibinfo  {journal}
				{{Quantum}}\ }\textbf {\bibinfo {volume} {4}},\ \bibinfo {pages} {376}
			(\bibinfo {year} {2020})}\BibitemShut {NoStop}%
		\bibitem [{\citenamefont {Malik}\ \emph {et~al.}(2016)\citenamefont {Malik},
			\citenamefont {Erhard}, \citenamefont {Huber}, \citenamefont {Krenn},
			\citenamefont {Fickler},\ and\ \citenamefont {Zeilinger}}]{malik16}%
		\BibitemOpen
		\bibfield  {author} {\bibinfo {author} {\bibfnamefont {M.}~\bibnamefont
				{Malik}}, \bibinfo {author} {\bibfnamefont {M.}~\bibnamefont {Erhard}},
			\bibinfo {author} {\bibfnamefont {M.}~\bibnamefont {Huber}}, \bibinfo
			{author} {\bibfnamefont {M.}~\bibnamefont {Krenn}}, \bibinfo {author}
			{\bibfnamefont {R.}~\bibnamefont {Fickler}}, \ and\ \bibinfo {author}
			{\bibfnamefont {A.}~\bibnamefont {Zeilinger}},\ }\href
		{http://dx.doi.org/10.1038/nphoton.2016.12} {\bibfield  {journal} {\bibinfo
				{journal} {Nat. Photon.}\ }\textbf {\bibinfo {volume} {10}},\ \bibinfo
			{pages} {248} (\bibinfo {year} {2016})}\BibitemShut {NoStop}%
		\bibitem [{\citenamefont {Zhang}\ \emph {et~al.}(2017)\citenamefont {Zhang},
			\citenamefont {Agnew}, \citenamefont {Roger}, \citenamefont {Roux},
			\citenamefont {Konrad}, \citenamefont {Faccio}, \citenamefont {Leach},\ and\
			\citenamefont {Forbes}}]{zhang17n}%
		\BibitemOpen
		\bibfield  {author} {\bibinfo {author} {\bibfnamefont {Y.}~\bibnamefont
				{Zhang}}, \bibinfo {author} {\bibfnamefont {M.}~\bibnamefont {Agnew}},
			\bibinfo {author} {\bibfnamefont {T.}~\bibnamefont {Roger}}, \bibinfo
			{author} {\bibfnamefont {F.~S.}\ \bibnamefont {Roux}}, \bibinfo {author}
			{\bibfnamefont {T.}~\bibnamefont {Konrad}}, \bibinfo {author} {\bibfnamefont
				{D.}~\bibnamefont {Faccio}}, \bibinfo {author} {\bibfnamefont
				{J.}~\bibnamefont {Leach}}, \ and\ \bibinfo {author} {\bibfnamefont
				{A.}~\bibnamefont {Forbes}},\ }\href
		{https://doi.org/10.1038/s41467-017-00706-1} {\bibfield  {journal} {\bibinfo
				{journal} {Nat. Commun.}\ }\textbf {\bibinfo {volume} {8}},\ \bibinfo {pages}
			{632} (\bibinfo {year} {2017})}\BibitemShut {NoStop}%
		\bibitem [{\citenamefont {Kues}\ \emph {et~al.}(2017)\citenamefont {Kues},
			\citenamefont {Reimer}, \citenamefont {Roztocki}, \citenamefont {Cort\'es},
			\citenamefont {Sciara}, \citenamefont {Wetzel}, \citenamefont {Zhang},
			\citenamefont {Cino}, \citenamefont {Chu}, \citenamefont {Little},
			\citenamefont {Moss}, \citenamefont {Caspani}, \citenamefont {Aza\~na},\ and\
			\citenamefont {Morandotti}}]{kues17}%
		\BibitemOpen
		\bibfield  {author} {\bibinfo {author} {\bibfnamefont {M.}~\bibnamefont
				{Kues}}, \bibinfo {author} {\bibfnamefont {C.}~\bibnamefont {Reimer}},
			\bibinfo {author} {\bibfnamefont {P.}~\bibnamefont {Roztocki}}, \bibinfo
			{author} {\bibfnamefont {L.~R.}\ \bibnamefont {Cort\'es}}, \bibinfo {author}
			{\bibfnamefont {S.}~\bibnamefont {Sciara}}, \bibinfo {author} {\bibfnamefont
				{B.}~\bibnamefont {Wetzel}}, \bibinfo {author} {\bibfnamefont
				{Y.}~\bibnamefont {Zhang}}, \bibinfo {author} {\bibfnamefont
				{A.}~\bibnamefont {Cino}}, \bibinfo {author} {\bibfnamefont {S.~T.}\
				\bibnamefont {Chu}}, \bibinfo {author} {\bibfnamefont {B.~E.}\ \bibnamefont
				{Little}}, \bibinfo {author} {\bibfnamefont {D.~J.}\ \bibnamefont {Moss}},
			\bibinfo {author} {\bibfnamefont {L.}~\bibnamefont {Caspani}}, \bibinfo
			{author} {\bibfnamefont {J.}~\bibnamefont {Aza\~na}}, \ and\ \bibinfo
			{author} {\bibfnamefont {R.}~\bibnamefont {Morandotti}},\ }\href
		{https://doi.org/10.1038/nature22986} {\bibfield  {journal} {\bibinfo
				{journal} {Nature}\ }\textbf {\bibinfo {volume} {546}},\ \bibinfo {pages}
			{622} (\bibinfo {year} {2017})}\BibitemShut {NoStop}%
		\bibitem [{\citenamefont {Erhard}\ \emph
			{et~al.}(2018{\natexlab{b}})\citenamefont {Erhard}, \citenamefont {Malik},
			\citenamefont {Krenn},\ and\ \citenamefont {Zeilinger}}]{erhard18}%
		\BibitemOpen
		\bibfield  {author} {\bibinfo {author} {\bibfnamefont {M.}~\bibnamefont
				{Erhard}}, \bibinfo {author} {\bibfnamefont {M.}~\bibnamefont {Malik}},
			\bibinfo {author} {\bibfnamefont {M.}~\bibnamefont {Krenn}}, \ and\ \bibinfo
			{author} {\bibfnamefont {A.}~\bibnamefont {Zeilinger}},\ }\href
		{https://doi.org/10.1038/s41566-018-0257-6} {\bibfield  {journal} {\bibinfo
				{journal} {Nat. Photon.}\ }\textbf {\bibinfo {volume} {12}},\ \bibinfo
			{pages} {759} (\bibinfo {year} {2018}{\natexlab{b}})}\BibitemShut {NoStop}%
		\bibitem [{\citenamefont {Wang}\ \emph {et~al.}(2018)\citenamefont {Wang},
			\citenamefont {Luo}, \citenamefont {Huang}, \citenamefont {Chen},
			\citenamefont {Su}, \citenamefont {Liu}, \citenamefont {Chen}, \citenamefont
			{Li}, \citenamefont {Fang}, \citenamefont {Jiang}, \citenamefont {Zhang},
			\citenamefont {Li}, \citenamefont {Liu}, \citenamefont {Lu},\ and\
			\citenamefont {Pan}}]{wang18}%
		\BibitemOpen
		\bibfield  {author} {\bibinfo {author} {\bibfnamefont {X.-L.}\ \bibnamefont
				{Wang}}, \bibinfo {author} {\bibfnamefont {Y.-H.}\ \bibnamefont {Luo}},
			\bibinfo {author} {\bibfnamefont {H.-L.}\ \bibnamefont {Huang}}, \bibinfo
			{author} {\bibfnamefont {M.-C.}\ \bibnamefont {Chen}}, \bibinfo {author}
			{\bibfnamefont {Z.-E.}\ \bibnamefont {Su}}, \bibinfo {author} {\bibfnamefont
				{C.}~\bibnamefont {Liu}}, \bibinfo {author} {\bibfnamefont {C.}~\bibnamefont
				{Chen}}, \bibinfo {author} {\bibfnamefont {W.}~\bibnamefont {Li}}, \bibinfo
			{author} {\bibfnamefont {Y.-Q.}\ \bibnamefont {Fang}}, \bibinfo {author}
			{\bibfnamefont {X.}~\bibnamefont {Jiang}}, \bibinfo {author} {\bibfnamefont
				{J.}~\bibnamefont {Zhang}}, \bibinfo {author} {\bibfnamefont
				{L.}~\bibnamefont {Li}}, \bibinfo {author} {\bibfnamefont {N.-L.}\
				\bibnamefont {Liu}}, \bibinfo {author} {\bibfnamefont {C.-Y.}\ \bibnamefont
				{Lu}}, \ and\ \bibinfo {author} {\bibfnamefont {J.-W.}\ \bibnamefont {Pan}},\
		}\href {\doibase 10.1103/PhysRevLett.120.260502} {\bibfield  {journal}
			{\bibinfo  {journal} {Phys. Rev. Lett.}\ }\textbf {\bibinfo {volume} {120}},\
			\bibinfo {pages} {260502} (\bibinfo {year} {2018})}\BibitemShut {NoStop}%
		\bibitem [{\citenamefont {Reimer}\ \emph {et~al.}(2019)\citenamefont {Reimer},
			\citenamefont {Sciara}, \citenamefont {Roztocki}, \citenamefont {Islam},
			\citenamefont {Romero~Cort\'es}, \citenamefont {Zhang}, \citenamefont
			{Fischer}, \citenamefont {Loranger}, \citenamefont {Kashyap}, \citenamefont
			{Cino}, \citenamefont {Chu}, \citenamefont {Little}, \citenamefont {Moss},
			\citenamefont {Caspani}, \citenamefont {Munro}, \citenamefont {Aza\~na},
			\citenamefont {Kues},\ and\ \citenamefont {Morandotti}}]{reimer19}%
		\BibitemOpen
		\bibfield  {author} {\bibinfo {author} {\bibfnamefont {C.}~\bibnamefont
				{Reimer}}, \bibinfo {author} {\bibfnamefont {S.}~\bibnamefont {Sciara}},
			\bibinfo {author} {\bibfnamefont {P.}~\bibnamefont {Roztocki}}, \bibinfo
			{author} {\bibfnamefont {M.}~\bibnamefont {Islam}}, \bibinfo {author}
			{\bibfnamefont {L.}~\bibnamefont {Romero~Cort\'es}}, \bibinfo {author}
			{\bibfnamefont {Y.}~\bibnamefont {Zhang}}, \bibinfo {author} {\bibfnamefont
				{B.}~\bibnamefont {Fischer}}, \bibinfo {author} {\bibfnamefont
				{S.}~\bibnamefont {Loranger}}, \bibinfo {author} {\bibfnamefont
				{R.}~\bibnamefont {Kashyap}}, \bibinfo {author} {\bibfnamefont
				{A.}~\bibnamefont {Cino}}, \bibinfo {author} {\bibfnamefont {S.~T.}\
				\bibnamefont {Chu}}, \bibinfo {author} {\bibfnamefont {B.~E.}\ \bibnamefont
				{Little}}, \bibinfo {author} {\bibfnamefont {D.~J.}\ \bibnamefont {Moss}},
			\bibinfo {author} {\bibfnamefont {L.}~\bibnamefont {Caspani}}, \bibinfo
			{author} {\bibfnamefont {W.~J.}\ \bibnamefont {Munro}}, \bibinfo {author}
			{\bibfnamefont {J.}~\bibnamefont {Aza\~na}}, \bibinfo {author} {\bibfnamefont
				{M.}~\bibnamefont {Kues}}, \ and\ \bibinfo {author} {\bibfnamefont
				{R.}~\bibnamefont {Morandotti}},\ }\href
		{https://doi.org/10.1038/s41567-018-0347-x} {\bibfield  {journal} {\bibinfo
				{journal} {Nat. Phys.}\ }\textbf {\bibinfo {volume} {15}},\ \bibinfo {pages}
			{148} (\bibinfo {year} {2019})}\BibitemShut {NoStop}%
		\bibitem [{\citenamefont {Tischler}\ \emph {et~al.}(2018)\citenamefont
			{Tischler}, \citenamefont {Ghafari}, \citenamefont {Baker}, \citenamefont
			{Slussarenko}, \citenamefont {Patel}, \citenamefont {Weston}, \citenamefont
			{Wollmann}, \citenamefont {Shalm}, \citenamefont {Verma}, \citenamefont
			{Nam}, \citenamefont {Nguyen}, \citenamefont {Wiseman},\ and\ \citenamefont
			{Pryde}}]{tischler18p}%
		\BibitemOpen
		\bibfield  {author} {\bibinfo {author} {\bibfnamefont {N.}~\bibnamefont
				{Tischler}}, \bibinfo {author} {\bibfnamefont {F.}~\bibnamefont {Ghafari}},
			\bibinfo {author} {\bibfnamefont {T.~J.}\ \bibnamefont {Baker}}, \bibinfo
			{author} {\bibfnamefont {S.}~\bibnamefont {Slussarenko}}, \bibinfo {author}
			{\bibfnamefont {R.~B.}\ \bibnamefont {Patel}}, \bibinfo {author}
			{\bibfnamefont {M.~M.}\ \bibnamefont {Weston}}, \bibinfo {author}
			{\bibfnamefont {S.}~\bibnamefont {Wollmann}}, \bibinfo {author}
			{\bibfnamefont {L.~K.}\ \bibnamefont {Shalm}}, \bibinfo {author}
			{\bibfnamefont {V.~B.}\ \bibnamefont {Verma}}, \bibinfo {author}
			{\bibfnamefont {S.~W.}\ \bibnamefont {Nam}}, \bibinfo {author} {\bibfnamefont
				{H.~C.}\ \bibnamefont {Nguyen}}, \bibinfo {author} {\bibfnamefont {H.~M.}\
				\bibnamefont {Wiseman}}, \ and\ \bibinfo {author} {\bibfnamefont {G.~J.}\
				\bibnamefont {Pryde}},\ }\href {\doibase 10.1103/PhysRevLett.121.100401}
		{\bibfield  {journal} {\bibinfo  {journal} {Phys. Rev. Lett.}\ }\textbf
			{\bibinfo {volume} {121}},\ \bibinfo {pages} {100401} (\bibinfo {year}
			{2018})}\BibitemShut {NoStop}%
		\bibitem [{\citenamefont {Saunders}\ \emph {et~al.}(2010)\citenamefont
			{Saunders}, \citenamefont {Jones}, \citenamefont {Wiseman},\ and\
			\citenamefont {Pryde}}]{saunders10}%
		\BibitemOpen
		\bibfield  {author} {\bibinfo {author} {\bibfnamefont {D.~J.}\ \bibnamefont
				{Saunders}}, \bibinfo {author} {\bibfnamefont {S.~J.}\ \bibnamefont {Jones}},
			\bibinfo {author} {\bibfnamefont {H.~M.}\ \bibnamefont {Wiseman}}, \ and\
			\bibinfo {author} {\bibfnamefont {G.~J.}\ \bibnamefont {Pryde}},\ }\href
		{http://dx.doi.org/10.1038/nphys1766} {\bibfield  {journal} {\bibinfo
				{journal} {Nat. Phys.}\ }\textbf {\bibinfo {volume} {6}},\ \bibinfo {pages}
			{845} (\bibinfo {year} {2010})}\BibitemShut {NoStop}%
		\bibitem [{\citenamefont {Kocsis}\ \emph {et~al.}(2015)\citenamefont {Kocsis},
			\citenamefont {Hall}, \citenamefont {Bennet}, \citenamefont {Saunders},\ and\
			\citenamefont {Pryde}}]{kocsis15}%
		\BibitemOpen
		\bibfield  {author} {\bibinfo {author} {\bibfnamefont {S.}~\bibnamefont
				{Kocsis}}, \bibinfo {author} {\bibfnamefont {M.~J.~W.}\ \bibnamefont {Hall}},
			\bibinfo {author} {\bibfnamefont {A.~J.}\ \bibnamefont {Bennet}}, \bibinfo
			{author} {\bibfnamefont {D.~J.}\ \bibnamefont {Saunders}}, \ and\ \bibinfo
			{author} {\bibfnamefont {G.~J.}\ \bibnamefont {Pryde}},\ }\href
		{http://dx.doi.org/10.1038/ncomms6886} {\bibfield  {journal} {\bibinfo
				{journal} {Nat. Commun.}\ }\textbf {\bibinfo {volume} {6}},\ \bibinfo {pages}
			{5886} (\bibinfo {year} {2015})}\BibitemShut {NoStop}%
		\bibitem [{\citenamefont {Passaro}\ \emph {et~al.}(2015)\citenamefont
			{Passaro}, \citenamefont {Cavalcanti}, \citenamefont {Skrzypczyk},\ and\
			\citenamefont {Ac{\'{\i}}n}}]{passaro15}%
		\BibitemOpen
		\bibfield  {author} {\bibinfo {author} {\bibfnamefont {E.}~\bibnamefont
				{Passaro}}, \bibinfo {author} {\bibfnamefont {D.}~\bibnamefont {Cavalcanti}},
			\bibinfo {author} {\bibfnamefont {P.}~\bibnamefont {Skrzypczyk}}, \ and\
			\bibinfo {author} {\bibfnamefont {A.}~\bibnamefont {Ac{\'{\i}}n}},\ }\href
		{\doibase 10.1088/1367-2630/17/11/113010} {\bibfield  {journal} {\bibinfo
				{journal} {New J. Phys.}\ }\textbf {\bibinfo {volume} {17}},\ \bibinfo
			{pages} {113010} (\bibinfo {year} {2015})}\BibitemShut {NoStop}%
		\bibitem [{\citenamefont {Zeng}\ \emph {et~al.}(2018)\citenamefont {Zeng},
			\citenamefont {Wang}, \citenamefont {Li},\ and\ \citenamefont
			{Zhang}}]{zeng18}%
		\BibitemOpen
		\bibfield  {author} {\bibinfo {author} {\bibfnamefont {Q.}~\bibnamefont
				{Zeng}}, \bibinfo {author} {\bibfnamefont {B.}~\bibnamefont {Wang}}, \bibinfo
			{author} {\bibfnamefont {P.}~\bibnamefont {Li}}, \ and\ \bibinfo {author}
			{\bibfnamefont {X.}~\bibnamefont {Zhang}},\ }\href {\doibase
			10.1103/PhysRevLett.120.030401} {\bibfield  {journal} {\bibinfo  {journal}
				{Phys. Rev. Lett.}\ }\textbf {\bibinfo {volume} {120}},\ \bibinfo {pages}
			{030401} (\bibinfo {year} {2018})}\BibitemShut {NoStop}%
		\bibitem [{\citenamefont {Dabrowski}\ \emph {et~al.}(2018)\citenamefont
			{Dabrowski}, \citenamefont {Mazelanik}, \citenamefont {Parniak},
			\citenamefont {Leszczy\ifmmode~\acute{n}\else \'{n}\fi{}ski}, \citenamefont
			{Lipka},\ and\ \citenamefont {Wasilewski}}]{dabrowski18}%
		\BibitemOpen
		\bibfield  {author} {\bibinfo {author} {\bibfnamefont {M.}~\bibnamefont
				{Dabrowski}}, \bibinfo {author} {\bibfnamefont {M.}~\bibnamefont
				{Mazelanik}}, \bibinfo {author} {\bibfnamefont {M.}~\bibnamefont {Parniak}},
			\bibinfo {author} {\bibfnamefont {A.}~\bibnamefont
				{Leszczy\ifmmode~\acute{n}\else \'{n}\fi{}ski}}, \bibinfo {author}
			{\bibfnamefont {M.}~\bibnamefont {Lipka}}, \ and\ \bibinfo {author}
			{\bibfnamefont {W.}~\bibnamefont {Wasilewski}},\ }\href {\doibase
			10.1103/PhysRevA.98.042126} {\bibfield  {journal} {\bibinfo  {journal} {Phys.
					Rev. A}\ }\textbf {\bibinfo {volume} {98}},\ \bibinfo {pages} {042126}
			(\bibinfo {year} {2018})}\BibitemShut {NoStop}%
		\bibitem [{\citenamefont {Designolle}\ \emph {et~al.}(2021)\citenamefont
			{Designolle}, \citenamefont {Srivastav}, \citenamefont {Uola}, \citenamefont
			{Valencia}, \citenamefont {McCutcheon}, \citenamefont {Malik},\ and\
			\citenamefont {Brunner}}]{designolle20}%
		\BibitemOpen
		\bibfield  {author} {\bibinfo {author} {\bibfnamefont {S.}~\bibnamefont
				{Designolle}}, \bibinfo {author} {\bibfnamefont {V.}~\bibnamefont
				{Srivastav}}, \bibinfo {author} {\bibfnamefont {R.}~\bibnamefont {Uola}},
			\bibinfo {author} {\bibfnamefont {N.~H.}\ \bibnamefont {Valencia}}, \bibinfo
			{author} {\bibfnamefont {W.}~\bibnamefont {McCutcheon}}, \bibinfo {author}
			{\bibfnamefont {M.}~\bibnamefont {Malik}}, \ and\ \bibinfo {author}
			{\bibfnamefont {N.}~\bibnamefont {Brunner}},\ }\href@noop {} {\bibfield
			{journal} {\bibinfo  {journal} {Phys. Rev. Lett}\ }\textbf {\bibinfo {volume}
				{126}},\ \bibinfo {pages} {200404} (\bibinfo {year} {2021})}\BibitemShut
		{NoStop}%
		\bibitem [{\citenamefont {Guerreiro}\ \emph {et~al.}(2016)\citenamefont
			{Guerreiro}, \citenamefont {Monteiro}, \citenamefont {Martin}, \citenamefont
			{Brask}, \citenamefont {V\'ertesi}, \citenamefont {Korzh}, \citenamefont
			{Caloz}, \citenamefont {Bussi\`eres}, \citenamefont {Verma}, \citenamefont
			{Lita}, \citenamefont {Mirin}, \citenamefont {Nam}, \citenamefont {Marsilli},
			\citenamefont {Shaw}, \citenamefont {Gisin}, \citenamefont {Brunner},
			\citenamefont {Zbinden},\ and\ \citenamefont {Thew}}]{guerreiro16}%
		\BibitemOpen
		\bibfield  {author} {\bibinfo {author} {\bibfnamefont {T.}~\bibnamefont
				{Guerreiro}}, \bibinfo {author} {\bibfnamefont {F.}~\bibnamefont {Monteiro}},
			\bibinfo {author} {\bibfnamefont {A.}~\bibnamefont {Martin}}, \bibinfo
			{author} {\bibfnamefont {J.~B.}\ \bibnamefont {Brask}}, \bibinfo {author}
			{\bibfnamefont {T.}~\bibnamefont {V\'ertesi}}, \bibinfo {author}
			{\bibfnamefont {B.}~\bibnamefont {Korzh}}, \bibinfo {author} {\bibfnamefont
				{M.}~\bibnamefont {Caloz}}, \bibinfo {author} {\bibfnamefont
				{F.}~\bibnamefont {Bussi\`eres}}, \bibinfo {author} {\bibfnamefont {V.~B.}\
				\bibnamefont {Verma}}, \bibinfo {author} {\bibfnamefont {A.~E.}\ \bibnamefont
				{Lita}}, \bibinfo {author} {\bibfnamefont {R.~P.}\ \bibnamefont {Mirin}},
			\bibinfo {author} {\bibfnamefont {S.~W.}\ \bibnamefont {Nam}}, \bibinfo
			{author} {\bibfnamefont {F.}~\bibnamefont {Marsilli}}, \bibinfo {author}
			{\bibfnamefont {M.~D.}\ \bibnamefont {Shaw}}, \bibinfo {author}
			{\bibfnamefont {N.}~\bibnamefont {Gisin}}, \bibinfo {author} {\bibfnamefont
				{N.}~\bibnamefont {Brunner}}, \bibinfo {author} {\bibfnamefont
				{H.}~\bibnamefont {Zbinden}}, \ and\ \bibinfo {author} {\bibfnamefont
				{R.~T.}\ \bibnamefont {Thew}},\ }\href {\doibase
			10.1103/PhysRevLett.117.070404} {\bibfield  {journal} {\bibinfo  {journal}
				{Phys. Rev. Lett.}\ }\textbf {\bibinfo {volume} {117}},\ \bibinfo {pages}
			{070404} (\bibinfo {year} {2016})}\BibitemShut {NoStop}%
		\bibitem [{\citenamefont {Weston}\ \emph {et~al.}(2016)\citenamefont {Weston},
			\citenamefont {Chrzanowski}, \citenamefont {Wollmann}, \citenamefont
			{Boston}, \citenamefont {Ho}, \citenamefont {Shalm}, \citenamefont {Verma},
			\citenamefont {Allman}, \citenamefont {Nam}, \citenamefont {Patel},
			\citenamefont {Slussarenko},\ and\ \citenamefont {Pryde}}]{weston16}%
		\BibitemOpen
		\bibfield  {author} {\bibinfo {author} {\bibfnamefont {M.~M.}\ \bibnamefont
				{Weston}}, \bibinfo {author} {\bibfnamefont {H.~M.}\ \bibnamefont
				{Chrzanowski}}, \bibinfo {author} {\bibfnamefont {S.}~\bibnamefont
				{Wollmann}}, \bibinfo {author} {\bibfnamefont {A.}~\bibnamefont {Boston}},
			\bibinfo {author} {\bibfnamefont {J.}~\bibnamefont {Ho}}, \bibinfo {author}
			{\bibfnamefont {L.~K.}\ \bibnamefont {Shalm}}, \bibinfo {author}
			{\bibfnamefont {V.~B.}\ \bibnamefont {Verma}}, \bibinfo {author}
			{\bibfnamefont {M.~S.}\ \bibnamefont {Allman}}, \bibinfo {author}
			{\bibfnamefont {S.~W.}\ \bibnamefont {Nam}}, \bibinfo {author} {\bibfnamefont
				{R.~B.}\ \bibnamefont {Patel}}, \bibinfo {author} {\bibfnamefont
				{S.}~\bibnamefont {Slussarenko}}, \ and\ \bibinfo {author} {\bibfnamefont
				{G.~J.}\ \bibnamefont {Pryde}},\ }\href {\doibase 10.1364/OE.24.010869}
		{\bibfield  {journal} {\bibinfo  {journal} {Opt. Express}\ }\textbf {\bibinfo
				{volume} {24}},\ \bibinfo {pages} {10869} (\bibinfo {year}
			{2016})}\BibitemShut {NoStop}%
		\bibitem [{\citenamefont {Evans}\ \emph {et~al.}(2010)\citenamefont {Evans},
			\citenamefont {Bennink}, \citenamefont {Grice}, \citenamefont {Humble},\ and\
			\citenamefont {Schaake}}]{evans10}%
		\BibitemOpen
		\bibfield  {author} {\bibinfo {author} {\bibfnamefont {P.~G.}\ \bibnamefont
				{Evans}}, \bibinfo {author} {\bibfnamefont {R.~S.}\ \bibnamefont {Bennink}},
			\bibinfo {author} {\bibfnamefont {W.~P.}\ \bibnamefont {Grice}}, \bibinfo
			{author} {\bibfnamefont {T.~S.}\ \bibnamefont {Humble}}, \ and\ \bibinfo
			{author} {\bibfnamefont {J.}~\bibnamefont {Schaake}},\ }\href {\doibase
			10.1103/PhysRevLett.105.253601} {\bibfield  {journal} {\bibinfo  {journal}
				{Phys. Rev. Lett.}\ }\textbf {\bibinfo {volume} {105}},\ \bibinfo {pages}
			{253601} (\bibinfo {year} {2010})}\BibitemShut {NoStop}%
		\bibitem [{\citenamefont {Marsili}\ \emph {et~al.}(2013)\citenamefont
			{Marsili}, \citenamefont {Verma}, \citenamefont {Stern}, \citenamefont
			{Harrington}, \citenamefont {Lita}, \citenamefont {Gerrits}, \citenamefont
			{Vayshenker}, \citenamefont {Baek}, \citenamefont {Shaw}, \citenamefont
			{Mirin},\ and\ \citenamefont {Nam}}]{marsili13}%
		\BibitemOpen
		\bibfield  {author} {\bibinfo {author} {\bibfnamefont {F.}~\bibnamefont
				{Marsili}}, \bibinfo {author} {\bibfnamefont {V.~B.}\ \bibnamefont {Verma}},
			\bibinfo {author} {\bibfnamefont {J.~A.}\ \bibnamefont {Stern}}, \bibinfo
			{author} {\bibfnamefont {S.}~\bibnamefont {Harrington}}, \bibinfo {author}
			{\bibfnamefont {A.~E.}\ \bibnamefont {Lita}}, \bibinfo {author}
			{\bibfnamefont {T.}~\bibnamefont {Gerrits}}, \bibinfo {author} {\bibfnamefont
				{I.}~\bibnamefont {Vayshenker}}, \bibinfo {author} {\bibfnamefont
				{B.}~\bibnamefont {Baek}}, \bibinfo {author} {\bibfnamefont {M.~D.}\
				\bibnamefont {Shaw}}, \bibinfo {author} {\bibfnamefont {R.~P.}\ \bibnamefont
				{Mirin}}, \ and\ \bibinfo {author} {\bibfnamefont {S.~W.}\ \bibnamefont
				{Nam}},\ }\href {http://dx.doi.org/10.1038/nphoton.2013.13} {\bibfield
			{journal} {\bibinfo  {journal} {Nat. Photon.}\ }\textbf {\bibinfo {volume}
				{7}},\ \bibinfo {pages} {210} (\bibinfo {year} {2013})}\BibitemShut {NoStop}%
		\bibitem [{\citenamefont {Klyshko}(1980)}]{klyshko80}%
		\BibitemOpen
		\bibfield  {author} {\bibinfo {author} {\bibfnamefont {D.~N.}\ \bibnamefont
				{Klyshko}},\ }\href {http://stacks.iop.org/0049-1748/10/i=9/a=A09} {\bibfield
			{journal} {\bibinfo  {journal} {Sov. J. Quantum Electron.}\ }\textbf
			{\bibinfo {volume} {10}},\ \bibinfo {pages} {1112} (\bibinfo {year}
			{1980})}\BibitemShut {NoStop}%
		\bibitem [{\citenamefont {Marrucci}\ \emph
			{et~al.}(2006{\natexlab{a}})\citenamefont {Marrucci}, \citenamefont {Manzo},\
			and\ \citenamefont {Paparo}}]{marrucci06}%
		\BibitemOpen
		\bibfield  {author} {\bibinfo {author} {\bibfnamefont {L.}~\bibnamefont
				{Marrucci}}, \bibinfo {author} {\bibfnamefont {C.}~\bibnamefont {Manzo}}, \
			and\ \bibinfo {author} {\bibfnamefont {D.}~\bibnamefont {Paparo}},\ }\href
		{dx.doi.org/10.1103/PhysRevLett.96.163905} {\bibfield  {journal} {\bibinfo
				{journal} {Phys. Rev. Lett.}\ }\textbf {\bibinfo {volume} {96}},\ \bibinfo
			{pages} {163905} (\bibinfo {year} {2006}{\natexlab{a}})}\BibitemShut
		{NoStop}%
		\bibitem [{\citenamefont {Slussarenko}\ \emph {et~al.}(2011)\citenamefont
			{Slussarenko}, \citenamefont {Murauski}, \citenamefont {Du}, \citenamefont
			{Chigrinov}, \citenamefont {Marrucci},\ and\ \citenamefont
			{Santamato}}]{slussarenko11}%
		\BibitemOpen
		\bibfield  {author} {\bibinfo {author} {\bibfnamefont {S.}~\bibnamefont
				{Slussarenko}}, \bibinfo {author} {\bibfnamefont {A.}~\bibnamefont
				{Murauski}}, \bibinfo {author} {\bibfnamefont {T.}~\bibnamefont {Du}},
			\bibinfo {author} {\bibfnamefont {V.}~\bibnamefont {Chigrinov}}, \bibinfo
			{author} {\bibfnamefont {L.}~\bibnamefont {Marrucci}}, \ and\ \bibinfo
			{author} {\bibfnamefont {E.}~\bibnamefont {Santamato}},\ }\href
		{https://doi.org/10.1364/OE.19.004085} {\bibfield  {journal} {\bibinfo
				{journal} {Opt. Express}\ }\textbf {\bibinfo {volume} {19}},\ \bibinfo
			{pages} {4085} (\bibinfo {year} {2011})}\BibitemShut {NoStop}%
		\bibitem [{\citenamefont {Marrucci}\ \emph {et~al.}(2011)\citenamefont
			{Marrucci}, \citenamefont {Karimi}, \citenamefont {Slussarenko},
			\citenamefont {Piccirillo}, \citenamefont {Santamato}, \citenamefont
			{Nagali},\ and\ \citenamefont {Sciarrino}}]{rev_marrucci11}%
		\BibitemOpen
		\bibfield  {author} {\bibinfo {author} {\bibfnamefont {L.}~\bibnamefont
				{Marrucci}}, \bibinfo {author} {\bibfnamefont {E.}~\bibnamefont {Karimi}},
			\bibinfo {author} {\bibfnamefont {S.}~\bibnamefont {Slussarenko}}, \bibinfo
			{author} {\bibfnamefont {B.}~\bibnamefont {Piccirillo}}, \bibinfo {author}
			{\bibfnamefont {E.}~\bibnamefont {Santamato}}, \bibinfo {author}
			{\bibfnamefont {E.}~\bibnamefont {Nagali}}, \ and\ \bibinfo {author}
			{\bibfnamefont {F.}~\bibnamefont {Sciarrino}},\ }\href
		{https://dx.doi.org/10.1088/2040-8978/13/6/064001} {\bibfield  {journal}
			{\bibinfo  {journal} {J. Opt.}\ }\textbf {\bibinfo {volume} {13}},\ \bibinfo
			{pages} {064001} (\bibinfo {year} {2011})}\BibitemShut {NoStop}%
		\bibitem [{\citenamefont {Rubano}\ \emph {et~al.}(2019)\citenamefont {Rubano},
			\citenamefont {Cardano}, \citenamefont {Piccirillo},\ and\ \citenamefont
			{Marrucci}}]{rev_rubano19}%
		\BibitemOpen
		\bibfield  {author} {\bibinfo {author} {\bibfnamefont {A.}~\bibnamefont
				{Rubano}}, \bibinfo {author} {\bibfnamefont {F.}~\bibnamefont {Cardano}},
			\bibinfo {author} {\bibfnamefont {B.}~\bibnamefont {Piccirillo}}, \ and\
			\bibinfo {author} {\bibfnamefont {L.}~\bibnamefont {Marrucci}},\ }\href
		{\doibase 10.1364/JOSAB.36.000D70} {\bibfield  {journal} {\bibinfo  {journal}
				{J. Opt. Soc. Am. B}\ }\textbf {\bibinfo {volume} {36}},\ \bibinfo {pages}
			{D70} (\bibinfo {year} {2019})}\BibitemShut {NoStop}%
		\bibitem [{\citenamefont {Bhandari}(1997)}]{bhandari97}%
		\BibitemOpen
		\bibfield  {author} {\bibinfo {author} {\bibfnamefont {R.}~\bibnamefont
				{Bhandari}},\ }\href {https://doi.org/10.1016/S0370-1573(96)00029-4}
		{\bibfield  {journal} {\bibinfo  {journal} {Phys. Rep.}\ }\textbf {\bibinfo
				{volume} {281}},\ \bibinfo {pages} {1} (\bibinfo {year} {1997})}\BibitemShut
		{NoStop}%
		\bibitem [{\citenamefont {Cardano}\ \emph {et~al.}(2012)\citenamefont
			{Cardano}, \citenamefont {Karimi}, \citenamefont {Slussarenko}, \citenamefont
			{Marrucci}, \citenamefont {de~Lisio},\ and\ \citenamefont
			{Santamato}}]{cardano12}%
		\BibitemOpen
		\bibfield  {author} {\bibinfo {author} {\bibfnamefont {F.}~\bibnamefont
				{Cardano}}, \bibinfo {author} {\bibfnamefont {E.}~\bibnamefont {Karimi}},
			\bibinfo {author} {\bibfnamefont {S.}~\bibnamefont {Slussarenko}}, \bibinfo
			{author} {\bibfnamefont {L.}~\bibnamefont {Marrucci}}, \bibinfo {author}
			{\bibfnamefont {C.}~\bibnamefont {de~Lisio}}, \ and\ \bibinfo {author}
			{\bibfnamefont {E.}~\bibnamefont {Santamato}},\ }\href {\doibase
			10.1364/AO.51.0000C1} {\bibfield  {journal} {\bibinfo  {journal} {Appl.
					Opt.}\ }\textbf {\bibinfo {volume} {51}},\ \bibinfo {pages} {C1} (\bibinfo
			{year} {2012})}\BibitemShut {NoStop}%
		\bibitem [{\citenamefont {Marrucci}\ \emph
			{et~al.}(2006{\natexlab{b}})\citenamefont {Marrucci}, \citenamefont {Manzo},\
			and\ \citenamefont {Paparo}}]{marrucci06a}%
		\BibitemOpen
		\bibfield  {author} {\bibinfo {author} {\bibfnamefont {L.}~\bibnamefont
				{Marrucci}}, \bibinfo {author} {\bibfnamefont {C.}~\bibnamefont {Manzo}}, \
			and\ \bibinfo {author} {\bibfnamefont {D.}~\bibnamefont {Paparo}},\ }\href
		{dx.doi.org/10.1063/1.2207993} {\bibfield  {journal} {\bibinfo  {journal}
				{Appl. Phys. Lett.}\ }\textbf {\bibinfo {volume} {88}},\ \bibinfo {pages}
			{221102} (\bibinfo {year} {2006}{\natexlab{b}})}\BibitemShut {NoStop}%
		\bibitem [{\citenamefont {Nersisyan}\ \emph {et~al.}(2009)\citenamefont
			{Nersisyan}, \citenamefont {Tabiryan}, \citenamefont {Steeves},\ and\
			\citenamefont {Kimball}}]{tabiryan09}%
		\BibitemOpen
		\bibfield  {author} {\bibinfo {author} {\bibfnamefont {S.}~\bibnamefont
				{Nersisyan}}, \bibinfo {author} {\bibfnamefont {N.}~\bibnamefont {Tabiryan}},
			\bibinfo {author} {\bibfnamefont {D.~M.}\ \bibnamefont {Steeves}}, \ and\
			\bibinfo {author} {\bibfnamefont {B.~R.}\ \bibnamefont {Kimball}},\ }\href
		{\doibase 10.1364/OE.17.011926} {\bibfield  {journal} {\bibinfo  {journal}
				{Opt. Express}\ }\textbf {\bibinfo {volume} {17}},\ \bibinfo {pages} {11926}
			(\bibinfo {year} {2009})}\BibitemShut {NoStop}%
		\bibitem [{\citenamefont {Bomzon}\ \emph {et~al.}(2001)\citenamefont {Bomzon},
			\citenamefont {Kleiner},\ and\ \citenamefont {Hasman}}]{bomzon01a}%
		\BibitemOpen
		\bibfield  {author} {\bibinfo {author} {\bibfnamefont {Z.}~\bibnamefont
				{Bomzon}}, \bibinfo {author} {\bibfnamefont {V.}~\bibnamefont {Kleiner}}, \
			and\ \bibinfo {author} {\bibfnamefont {E.}~\bibnamefont {Hasman}},\ }\href
		{http://ol.osa.org/abstract.cfm?URI=ol-26-18-1424} {\bibfield  {journal}
			{\bibinfo  {journal} {Opt. Lett.}\ }\textbf {\bibinfo {volume} {26}},\
			\bibinfo {pages} {1424} (\bibinfo {year} {2001})}\BibitemShut {NoStop}%
		\bibitem [{\citenamefont {Karimi}\ \emph {et~al.}(2014)\citenamefont {Karimi},
			\citenamefont {Schulz}, \citenamefont {De~Leon}, \citenamefont {Qassim},
			\citenamefont {Upham},\ and\ \citenamefont {Boyd}}]{karimi14}%
		\BibitemOpen
		\bibfield  {author} {\bibinfo {author} {\bibfnamefont {E.}~\bibnamefont
				{Karimi}}, \bibinfo {author} {\bibfnamefont {S.~A.}\ \bibnamefont {Schulz}},
			\bibinfo {author} {\bibfnamefont {I.}~\bibnamefont {De~Leon}}, \bibinfo
			{author} {\bibfnamefont {H.}~\bibnamefont {Qassim}}, \bibinfo {author}
			{\bibfnamefont {J.}~\bibnamefont {Upham}}, \ and\ \bibinfo {author}
			{\bibfnamefont {R.~W.}\ \bibnamefont {Boyd}},\ }\href
		{https://doi.org/10.1038/lsa.2014.48} {\bibfield  {journal} {\bibinfo
				{journal} {Light Sci. Appl.}\ }\textbf {\bibinfo {volume} {3}},\ \bibinfo
			{pages} {e167} (\bibinfo {year} {2014})}\BibitemShut {NoStop}%
		\bibitem [{\citenamefont {Piccardo}\ and\ \citenamefont
			{Ambrosio}(2020)}]{rev_piccardo20}%
		\BibitemOpen
		\bibfield  {author} {\bibinfo {author} {\bibfnamefont {M.}~\bibnamefont
				{Piccardo}}\ and\ \bibinfo {author} {\bibfnamefont {A.}~\bibnamefont
				{Ambrosio}},\ }\href {\doibase 10.1063/5.0023338} {\bibfield  {journal}
			{\bibinfo  {journal} {Applied Physics Letters}\ }\textbf {\bibinfo {volume}
				{117}},\ \bibinfo {pages} {140501} (\bibinfo {year} {2020})}\BibitemShut
		{NoStop}%
		\bibitem [{\citenamefont {Cozzolino}\ \emph {et~al.}(2019)\citenamefont
			{Cozzolino}, \citenamefont {Polino}, \citenamefont {Valeri}, \citenamefont
			{Carvacho}, \citenamefont {Bacco}, \citenamefont {Spagnolo}, \citenamefont
			{Oxenl{\o}we},\ and\ \citenamefont {Sciarrino}}]{cozzolino19}%
		\BibitemOpen
		\bibfield  {author} {\bibinfo {author} {\bibfnamefont {D.}~\bibnamefont
				{Cozzolino}}, \bibinfo {author} {\bibfnamefont {E.}~\bibnamefont {Polino}},
			\bibinfo {author} {\bibfnamefont {M.}~\bibnamefont {Valeri}}, \bibinfo
			{author} {\bibfnamefont {G.}~\bibnamefont {Carvacho}}, \bibinfo {author}
			{\bibfnamefont {D.}~\bibnamefont {Bacco}}, \bibinfo {author} {\bibfnamefont
				{N.}~\bibnamefont {Spagnolo}}, \bibinfo {author} {\bibfnamefont {L.~K.~K.}\
				\bibnamefont {Oxenl{\o}we}}, \ and\ \bibinfo {author} {\bibfnamefont
				{F.}~\bibnamefont {Sciarrino}},\ }\href {\doibase 10.1117/1.AP.1.4.046005}
		{\bibfield  {journal} {\bibinfo  {journal} {Adv. Photonics}\ }\textbf
			{\bibinfo {volume} {1}},\ \bibinfo {pages} {1 } (\bibinfo {year}
			{2019})}\BibitemShut {NoStop}%
		\bibitem [{\citenamefont {James}\ \emph {et~al.}(2001)\citenamefont {James},
			\citenamefont {Kwiat}, \citenamefont {Munro},\ and\ \citenamefont
			{White}}]{james01}%
		\BibitemOpen
		\bibfield  {author} {\bibinfo {author} {\bibfnamefont {D.~F.~V.}\
				\bibnamefont {James}}, \bibinfo {author} {\bibfnamefont {P.~G.}\ \bibnamefont
				{Kwiat}}, \bibinfo {author} {\bibfnamefont {W.~J.}\ \bibnamefont {Munro}}, \
			and\ \bibinfo {author} {\bibfnamefont {A.~G.}\ \bibnamefont {White}},\ }\href
		{\doibase http://dx.doi.org/10.1103/PhysRevA.64.052312} {\bibfield  {journal}
			{\bibinfo  {journal} {Phys. Rev. A}\ }\textbf {\bibinfo {volume} {64}},\
			\bibinfo {pages} {052312} (\bibinfo {year} {2001})}\BibitemShut {NoStop}%
		\bibitem [{\citenamefont {Palsson}\ \emph {et~al.}(2012)\citenamefont
			{Palsson}, \citenamefont {Wallman}, \citenamefont {Bennet},\ and\
			\citenamefont {Pryde}}]{palsson12}%
		\BibitemOpen
		\bibfield  {author} {\bibinfo {author} {\bibfnamefont {M.~S.}\ \bibnamefont
				{Palsson}}, \bibinfo {author} {\bibfnamefont {J.~J.}\ \bibnamefont
				{Wallman}}, \bibinfo {author} {\bibfnamefont {A.~J.}\ \bibnamefont {Bennet}},
			\ and\ \bibinfo {author} {\bibfnamefont {G.~J.}\ \bibnamefont {Pryde}},\
		}\href {\doibase 10.1103/PhysRevA.86.032322} {\bibfield  {journal} {\bibinfo
				{journal} {Phys. Rev. A}\ }\textbf {\bibinfo {volume} {86}},\ \bibinfo
			{pages} {032322} (\bibinfo {year} {2012})}\BibitemShut {NoStop}%
		\bibitem [{\citenamefont {Shadbolt}\ \emph {et~al.}(2012)\citenamefont
			{Shadbolt}, \citenamefont {V\'{e}rtesi}, \citenamefont {Liang}, \citenamefont
			{Branciard}, \citenamefont {Brunner},\ and\ \citenamefont
			{O'Brien}}]{shadbolt12}%
		\BibitemOpen
		\bibfield  {author} {\bibinfo {author} {\bibfnamefont {P.}~\bibnamefont
				{Shadbolt}}, \bibinfo {author} {\bibfnamefont {T.}~\bibnamefont
				{V\'{e}rtesi}}, \bibinfo {author} {\bibfnamefont {Y.-C.}\ \bibnamefont
				{Liang}}, \bibinfo {author} {\bibfnamefont {C.}~\bibnamefont {Branciard}},
			\bibinfo {author} {\bibfnamefont {N.}~\bibnamefont {Brunner}}, \ and\
			\bibinfo {author} {\bibfnamefont {J.~L.}\ \bibnamefont {O'Brien}},\ }\href
		{https://doi.org/10.1038/srep00470} {\bibfield  {journal} {\bibinfo
				{journal} {Sci. Rep.}\ }\textbf {\bibinfo {volume} {2}},\ \bibinfo {pages}
			{470} (\bibinfo {year} {2012})}\BibitemShut {NoStop}%
		\bibitem [{\citenamefont {Wollmann}\ \emph {et~al.}(2018)\citenamefont
			{Wollmann}, \citenamefont {Hall}, \citenamefont {Patel}, \citenamefont
			{Wiseman},\ and\ \citenamefont {Pryde}}]{wollmann18}%
		\BibitemOpen
		\bibfield  {author} {\bibinfo {author} {\bibfnamefont {S.}~\bibnamefont
				{Wollmann}}, \bibinfo {author} {\bibfnamefont {M.~J.~W.}\ \bibnamefont
				{Hall}}, \bibinfo {author} {\bibfnamefont {R.~B.}\ \bibnamefont {Patel}},
			\bibinfo {author} {\bibfnamefont {H.~M.}\ \bibnamefont {Wiseman}}, \ and\
			\bibinfo {author} {\bibfnamefont {G.~J.}\ \bibnamefont {Pryde}},\ }\href
		{\doibase 10.1103/PhysRevA.98.022333} {\bibfield  {journal} {\bibinfo
				{journal} {Phys. Rev. A}\ }\textbf {\bibinfo {volume} {98}},\ \bibinfo
			{pages} {022333} (\bibinfo {year} {2018})}\BibitemShut {NoStop}%
		\bibitem [{\citenamefont {Weston}\ \emph {et~al.}(2018)\citenamefont {Weston},
			\citenamefont {Slussarenko}, \citenamefont {Chrzanowski}, \citenamefont
			{Wollmann}, \citenamefont {Shalm}, \citenamefont {Verma}, \citenamefont
			{Allman}, \citenamefont {Nam},\ and\ \citenamefont {Pryde}}]{weston18}%
		\BibitemOpen
		\bibfield  {author} {\bibinfo {author} {\bibfnamefont {M.~M.}\ \bibnamefont
				{Weston}}, \bibinfo {author} {\bibfnamefont {S.}~\bibnamefont {Slussarenko}},
			\bibinfo {author} {\bibfnamefont {H.~M.}\ \bibnamefont {Chrzanowski}},
			\bibinfo {author} {\bibfnamefont {S.}~\bibnamefont {Wollmann}}, \bibinfo
			{author} {\bibfnamefont {L.~K.}\ \bibnamefont {Shalm}}, \bibinfo {author}
			{\bibfnamefont {V.~B.}\ \bibnamefont {Verma}}, \bibinfo {author}
			{\bibfnamefont {M.~S.}\ \bibnamefont {Allman}}, \bibinfo {author}
			{\bibfnamefont {S.~W.}\ \bibnamefont {Nam}}, \ and\ \bibinfo {author}
			{\bibfnamefont {G.~J.}\ \bibnamefont {Pryde}},\ }\href
		{http://advances.sciencemag.org/content/4/1/e1701230} {\bibfield  {journal}
			{\bibinfo  {journal} {Sci. Adv.}\ }\textbf {\bibinfo {volume} {4}},\ \bibinfo
			{pages} {e1701230} (\bibinfo {year} {2018})}\BibitemShut {NoStop}%
		\bibitem [{\citenamefont {Tsujimoto}\ \emph {et~al.}(2020)\citenamefont
			{Tsujimoto}, \citenamefont {You}, \citenamefont {Wakui}, \citenamefont
			{Fujiwara}, \citenamefont {Hayasaka}, \citenamefont {Miki}, \citenamefont
			{Terai}, \citenamefont {Sasaki}, \citenamefont {Dowling},\ and\ \citenamefont
			{Takeoka}}]{tsujimoto20}%
		\BibitemOpen
		\bibfield  {author} {\bibinfo {author} {\bibfnamefont {Y.}~\bibnamefont
				{Tsujimoto}}, \bibinfo {author} {\bibfnamefont {C.}~\bibnamefont {You}},
			\bibinfo {author} {\bibfnamefont {K.}~\bibnamefont {Wakui}}, \bibinfo
			{author} {\bibfnamefont {M.}~\bibnamefont {Fujiwara}}, \bibinfo {author}
			{\bibfnamefont {K.}~\bibnamefont {Hayasaka}}, \bibinfo {author}
			{\bibfnamefont {S.}~\bibnamefont {Miki}}, \bibinfo {author} {\bibfnamefont
				{H.}~\bibnamefont {Terai}}, \bibinfo {author} {\bibfnamefont
				{M.}~\bibnamefont {Sasaki}}, \bibinfo {author} {\bibfnamefont {J.~P.}\
				\bibnamefont {Dowling}}, \ and\ \bibinfo {author} {\bibfnamefont
				{M.}~\bibnamefont {Takeoka}},\ }\href {\doibase 10.1088/1367-2630/ab61da}
		{\bibfield  {journal} {\bibinfo  {journal} {New J. Phys.}\ }\textbf {\bibinfo
				{volume} {22}},\ \bibinfo {pages} {023008} (\bibinfo {year}
			{2020})}\BibitemShut {NoStop}%
		\bibitem [{\citenamefont {Milione}\ \emph {et~al.}(2015)\citenamefont
			{Milione}, \citenamefont {Nguyen}, \citenamefont {Leach}, \citenamefont
			{Nolan},\ and\ \citenamefont {Alfano}}]{milione15ol}%
		\BibitemOpen
		\bibfield  {author} {\bibinfo {author} {\bibfnamefont {G.}~\bibnamefont
				{Milione}}, \bibinfo {author} {\bibfnamefont {T.~A.}\ \bibnamefont {Nguyen}},
			\bibinfo {author} {\bibfnamefont {J.}~\bibnamefont {Leach}}, \bibinfo
			{author} {\bibfnamefont {D.~A.}\ \bibnamefont {Nolan}}, \ and\ \bibinfo
			{author} {\bibfnamefont {R.~R.}\ \bibnamefont {Alfano}},\ }\href {\doibase
			10.1364/OL.40.004887} {\bibfield  {journal} {\bibinfo  {journal} {Opt.
					Lett.}\ }\textbf {\bibinfo {volume} {40}},\ \bibinfo {pages} {4887} (\bibinfo
			{year} {2015})}\BibitemShut {NoStop}%
	\end{thebibliography}

\end{document}